\documentclass[conference]{IEEEtran}
\IEEEoverridecommandlockouts
\pdfoutput=1
\usepackage{cite}
\usepackage{amsmath,amssymb,amsfonts}

\newcommand{\ceil}[1]{\left\lceil #1 \right\rceil}
\usepackage{algorithmic}
\usepackage{graphicx}
\usepackage{textcomp}
\usepackage{xcolor}
\usepackage{float}
\usepackage{dsfont}
\usepackage{tabularx}
\usepackage{booktabs}
\usepackage{xcolor}
\usepackage[labelfont=bf,format=plain,justification=raggedright,singlelinecheck=false]{caption}

\usepackage{caption}
\usepackage{subcaption}
\usepackage[hyphens]{url}
\usepackage{hyperref}
\usepackage[english]{babel}
\usepackage{amsthm}
\usepackage{float}
\floatstyle{plaintop}
\restylefloat{table}

\theoremstyle{definition}
\newtheorem{definition}{Definition}

\theoremstyle{definition}
\newtheorem{lemma}{Lemma}

\usepackage{algorithm}
\usepackage{algorithmic}

\newcommand{\NAME}{\texttt{TxAllo}}
\newcommand{\globalNAME}{\texttt{G-TxAllo}}
\newcommand{\adaNAME}{\texttt{A-TxAllo}}
\newcommand{\capacity}{\lambda}
\newcommand{\ledger}{\mathcal{L}}
\newcommand{\Block}{\mathcal{B}}
\newcommand{\setnode}{\mathrm{V}}
\newcommand{\Tx}{\mathtt{Tx}}
\newcommand{\Inputacc}{\mathcal{A}^{in}_{\Tx}}
\newcommand{\outputacc}{\mathcal{A}^{out}_{\Tx}}

\newcommand{\shardacc}{\mathcal{A}}

\newcommand{\transactioninshard}{\mathcal{T}}
\newcommand{\allocation}{\phi}

\newcommand{\numshard}{k}
\newcommand{\Difficulty}{\eta}
\newcommand{\crossratio}{\gamma}
\newcommand{\balancemetrics}{\rho}
\newcommand{\workload}{\sigma}
\newcommand{\throughput}{\Lambda}
\newcommand{\graph}{\mathrm{G}}
\newcommand{\setaccount}{\mathcal{A}}
\newcommand{\setalltx}{\mathcal{T}}

\newcommand{\setcross}{\mathcal{T}^\mathcal{C}}
\newcommand{\setintra}{\mathcal{T}^\mathcal{I}}
\newcommand{\workloadaverage}{\bar{\workload}}
\newcommand{\Transactionacc}{\mathcal{A}_{\Tx}}
\newcommand{\setedgetx}{\mathcal{T}}
\newcommand{\latency}{\zeta}
\newcommand{\numrelashards}{\mu}
\newcommand{\convergence}{\varepsilon}
\newcommand{\candidate}{\mathds{C}}
\newcommand{\updateglo}{\tau_2}
\newcommand{\updateada}{\tau_1}
\newcommand{\adaV}{\hat{V}}
\newcommand{\parameters}{\theta}

\def\BibTeX{{\rm B\kern-.05em{\sc i\kern-.025em b}\kern-.08em
    T\kern-.1667em\lower.7ex\hbox{E}\kern-.125emX}}
\begin{document}

\title{\NAME: Dynamic Transaction Allocation in Sharded Blockchain Systems}
\author{\IEEEauthorblockN{Yuanzhe Zhang}
\IEEEauthorblockA{\textit{Monash University}\\
Melbourne, Australia \\
Yuanzhe.Zhang@monash.edu}
\and

\IEEEauthorblockN{Shirui Pan}
\IEEEauthorblockA{\textit{Griffith University}\\
Gold Coast, Australia \\
s.pan@griffith.edu.au}
\and

\IEEEauthorblockN{Jiangshan Yu$^{\ast}$}\thanks{*Corresponding author}
\IEEEauthorblockA{\textit{Monash University}\\
Melbourne, Australia \\
Jiangshan.Yu@monash.edu}

}

\maketitle

\begin{abstract}
The scalability problem has been one of the most significant barriers
  limiting the adoption of blockchains. 
  Blockchain sharding is a promising
  approach to this problem. 
  However, the sharding mechanism
  introduces a significant number of cross-shard
  transactions, which are expensive to process.
  
  This paper focuses on the transaction allocation problem to reduce the number of cross-shard transactions for better scalability. 
  In particular, we systematically formulate the transaction allocation
  problem and convert it to the community detection problem on a
  graph. 
 A deterministic and fast allocation scheme $\NAME{}$ is proposed to dynamically infer the allocation
  of accounts and their associated transactions. 
  It directly optimizes the system throughput, considering both the number of cross-shard transactions and the workload balance among shards.

  We evaluate the performance of $\NAME{}$ on an Ethereum dataset
  containing over 91 million transactions. Our evaluation results show
  that for a blockchain with 60 shards, $\NAME{}$ reduces the
  cross-shard transaction ratio from 98\% (by using traditional hash-based
  allocation) to about 12\%.
  In the meantime, the workload balance is well maintained.
  Compared with other methods, the execution time of
  $\NAME{}$ is almost negligible. For example, when updating the
  allocation every hour, the execution of $\NAME{}$ only takes 0.5
  seconds on average, whereas other concurrent works, such as
  BrokerChain (INFOCOM'22) leveraging the classic METIS method,
  require 422 seconds.

\end{abstract}

\section{Introduction}
Blockchains have become a disruptive technology that is transforming
the world. According to
CoinMarketCap\footnote{ \url{https://coinmarketcap.com}, data fetched on 20-Oct-2021.}, the total value of all listed cryptocurrencies is
over \$2.5 trillion USD, with a daily trading volume of over
\$91 billion USD. Blockchains can be divided into two categories: permissionless and permissioned blockchains. Permissionless blockchains are open to everyone to participate (e.g., Bitcoin~\cite{nakamoto2008bitcoin} and Ethereum~\cite{wood2014ethereum}), whereas permissioned blockchains have specific criteria to select a set of known and identified participants (e.g. Hyperledger Fabric~\cite{fabric}). 
This work focuses on permissionless blockchains.

A permissionless blockchain typically has two main types of participants: miners and clients. 
Clients create transactions and broadcast them to a peer-to-peer network 
and miners process them by executing a consensus algorithm to agree on totally-ordered transactions as an immutable ledger. 
However, due to the nature of decentralization, scalability has been a big issue for large-scale applications~\cite{kim2018survey,xie2019survey,eklund2019factors,chauhan2018blockchain,wang2019sok,yang2020review}.
Loosely speaking, we consider a blockchain system to be scalable if its throughput increases linearly
with the number of miners.
Taking the top 2 blockchain systems as an example, no matter how many miners are involved, Bitcoin and Ethereum can only process approximately 7 and 14 transactions per second, respectively.

To scale blockchains, the sharding technique has been proposed to separate and distribute miners and transactions into many parts (called shards) for concurrent processing~\cite{HanYZ20}. 
In this way, each shard is only required to process a fraction of transactions.
However, existing sharding protocols~\cite{luu2016secure,kokoris2018omniledger,zamani2018rapidchain,al2017chainspace,wang2019monoxide} rely on a hash-based method to allocate transactions into different shards.
As historical transaction patterns are not leveraged, this transaction allocation method introduces approximately 90\% cross-shard transactions, making it impractical in real-world systems~\cite{wang2019sok}.

A cross-shard transaction occurs when the processing requires the cooperation and consensus of multiple shards.
It requires an extra round of consensus, and different shards also need to synchronize their knowledge before the processing~\cite{han2021security}. 
This not only increases the latency of finalizing transactions but also
reduces the effectiveness of concurrency control, leading to limited
improvements in scalability.

To reduce the number of cross-shard transactions, recent works~\cite{fynn2018challenges, mizrahi2020blockchain, huang2022brokerchain} utilized a classic graph-based method, METIS~\cite{karypis1997metis}, to cluster frequently interacted accounts into one shard.
However, the METIS method considers the weight balance instead of the workload balance among shards.
Workload balance among shards is vital to sharding, resulting in better throughput, lower confirmation latency and order fairness of transactions~\cite{kelkar2020order}. 
In addition, fast execution is required as an open challenge for the periodical updating of the new allocation. 

In this work, we formalize the transaction allocation problem and convert it to the community detection problem on a graph, leveraging historical transaction patterns. 
We propose \NAME{}, an effective and efficient transaction allocation mechanism to reduce cross-shard transactions so that it improves the throughput of sharded blockchains. 
\NAME{} detects the community structure by directly optimizing the system throughput on the graph, considering both the cross-shard transaction ratio and the workload balance. 
It consists of two components: global algorithm \globalNAME{} and adaptive algorithm \adaNAME.
The former takes all historical transactions as input to obtain the precise allocation.
The latter takes the previous allocation results and the newly committed transaction as input to avoid the slowdown caused by the accumulation of transactions.

 \NAME{} is implemented on real-world Ethereum data with over 91 million transactions and over 12 million accounts. 
We illustrate that, for an arbitrary given number of shards,  \NAME{} reduces the cross-shard transactions in scale and keeps the balanced workload simultaneously, resulting in significant throughput improvement from existing works. 
For example, within 60 shards, \NAME{} can reduce the cross-shard transaction ratio to 12\%. In contrast, it is 98\% and 28\% by using the traditional hash-based method and the METIS method, respectively.
Furthermore, we show that the running time of \globalNAME{} and \adaNAME{} is 122 seconds and 0.55 seconds, respectively, which is significantly improved from the 422 seconds by the METIS method.

The main \textbf{contributions} of this work are: 
\begin{enumerate}
 \item We systematically formulate the transaction allocation problem in sharded blockchains and convert it to the community detection problem on graphs. In particular, 
 we first define the key concepts of a sharded blockchain, including the cross-shard transaction, the workload in a shard, the system throughput and the confirmation latency. Then, we construct a transaction graph to capture the historical transaction patterns and convert these key concepts to the context of this transaction graph.
 \item We propose a dynamic allocation mechanism \NAME{} to directly optimize the throughput on the transaction graph, which simultaneously considers workload balance and cross-shard transaction ratio. It is deterministic, adaptive and fast for execution, satisfying the specific requirements in blockchains. 
 \item We implement \NAME{} on real-world Ethereum data and achieve significant improvement in both performance and running time. We evaluate and compare \NAME{} with state-of-the-art baselines from various aspects, including cross-shard transaction ratio, workload balance, system throughput, transaction confirmation latency and running time.
\end{enumerate} 

The rest of this work is organized as follows. 
Section \ref{sec_background} introduces the background and related works. 
Section \ref{sec_for} formulates the transaction allocation problem and converts it to community detection on the graph. 
Section \ref{sec_design}
introduces the design challenges in
real-world sharded blockchains. 
Section \ref{sec_method} introduces our
proposed \NAME{} for transaction allocation. 
Section \ref{sec_exp} discusses the experimental results and the comparison with baselines.
Section~\ref{sec_integration} discusses integrating \NAME{} into existing sharding protocols.
Finally, we conclude this work in Section \ref{sec_conclu}.

\section{Background and related works}
\label{sec_background}

\subsection{Blockchains}
Blockchains can be UTXO-based (Unspent Transaction Output) or account-based in terms of the data model.
For example, Bitcoin utilizes the former and Ethereum utilizes the latter.
This work focuses on account-based blockchains, where the account is persistent and can be repeatedly used in the system. This nature preserves transaction patterns in the blockchain.

In account-based blockchains, a transaction modifies the states of associated input and output accounts, which indicate where the token is transferred from and to, respectively.
A transaction is allowed to include multiple inputs and outputs, which is called a multi-input multi-output transaction.
The accounts owned by ordinary clients are called Externally Owned Accounts (EOA).
Smart contracts in Ethereum can also have their own accounts called Contract Accounts (CA). Usually, a CA is more active than an EOA, as smart contracts can frequently call it. 

\subsection{Sharding in Blockchains}
The basic idea of sharding is to horizontally divide a database into small parts (called shards), originating from the traditional database domain~\cite{liu2012research}. 
Many works adopted this concept into both permissionless~\cite{luu2016secure, al2017chainspace, kokoris2018omniledger,zamani2018rapidchain, wang2019monoxide} and permissioned blockchains~\cite{dang2019towards, amiri2021sharper, rahnama2021ringbft} for the scalability problem. 
They allocate miners into different shards and each shard only processes a fraction of the total transactions, which is allocated in this shard. In this way, the expected scalability is achieved that the throughput increases linearly with the number of shards.

Elastico~\cite{luu2016secure} is a pioneering work to leverage sharding in the blockchain domain. To prevent single-shard take-over attacks from malicious miners, Elastico randomly allocates miners into different shards and reshuffles miners periodically in the reconfiguration phase. Other permissionless sharding protocols also utilize this periodical reshuffling process for security concerns~\cite{wang2019sok}.

However, cross-shard transactions are a significant barrier limiting the application of sharding protocols, which is not addressed in Elastico. In account-based blockchains, a cross-shard transaction occurs if multiple shards maintain this transaction's input and output accounts. 
Atomicity must be guaranteed that a transaction is either fully committed or fully aborted by all involved shards. This introduces additional communication overhead.

The following works of sharded blockchains focus on the security and efficiency of processing cross-shard transactions. 
Chainspace~\cite{al2017chainspace} proposes a BFT-based consensus to tackle the atomicity and consistency of cross-shard transactions, especially for smart contract platforms. OmniLedger~\cite{kokoris2018omniledger} proposes a two-phase client-driven consensus, Atomix, for cross-shard transactions. RapidChain~\cite{zamani2018rapidchain} develops an efficient routing mechanism for shards to discover each other with only a logarithmic latency and storage. Monoxide~\cite{wang2019monoxide} proposes Chu-ko-nu mining for the problem of mining power decreasing in sharding. AHL~\cite{dang2019towards} leverages a trusted execution environment (TEE) to improve the efficiency and fault tolerance of the BFT-based consensus in each shard. SharPer~\cite{amiri2021sharper} proposes to cluster miners by geographical distribution for concurrent processing. RingBFT~\cite{rahnama2021ringbft} proposes a ring topology for shards to achieve linear communication between them.

\subsection{Transaction Allocation}

Reducing the number of cross-shard transactions is an orthogonal direction for better scalability, in addition to the efficiency problem of processing each individual cross-shard transaction.
The protocols mentioned above randomly allocate accounts and transactions by a hash-based method, resulting in more than 90\% cross-shard transactions.
In particular, the account-based blockchain Chainspace~\cite{al2017chainspace} allocates accounts into a shard by the hash value of their addresses, i.e.,
SHA256 (address) mod $k$, where $k$ is the number of shards.
Based on this account allocation of the inputs and outputs, transactions are allocated into the affected shards to process.
Monoxide~\cite{wang2019monoxide} allocates accounts into $2^k$ shards by the first $k$ bits of the hash value of their addresses.
OmniLedger~\cite{kokoris2018omniledger} and RapidChain~\cite{zamani2018rapidchain}, leveraging the UTXO-based model, both randomly allocate UTXOs based on the first several bits of their hash value, similar to Monoxide.

Recent works have explored the transaction allocation problem by leveraging historical transaction patterns.
For UTXO-based blockchains, Nguyen et al. proposed Optchain~\cite{nguyen2019optchain} to model each transaction as a node in a graph, building a lightweight solution with temporal balance among shards.
For account-based blockchains, the proposed methods can be graph-based or individual transaction-level.
Graph-based methods aim to globally derive a partition based on historical transactions, whereas individual transaction-level methods determine whether to move affected accounts to another shard when each transaction comes.
Fynn et al. first identified the transaction allocation problem and modelled historical transaction patterns as a graph~\cite{fynn2018challenges}. 
They utilize the METIS method~\cite{karypis1997metis} to obtain the partition of accounts and associated transactions.
Avi et al. further considered the memory-efficient problem of the allocation mapping~\cite{mizrahi2020blockchain}.
Huang et al. proposed BrokerChain~\cite{huang2022brokerchain} to further address the workload balance problem by introducing a special kind of participants called brokers on top of the transaction allocation problem. 
However, the backbone allocation algorithms of all these graph-based methods are the classic graph partition method, METIS.

The METIS method is a series of multi-level methods that has achieved massive success in the graph partition domain. Traditional sharded databases, such as Schism~\cite{curino2010schism}, also utilize the METIS method for partitioning.
This method relies on three phases: coarsening, partitioning and refinement, to achieve both low inter-cluster weights and balanced weights in each cluster.
However, the METIS method is not specifically designed for our problem in blockchains.
First, the balance is in terms of weights rather than the workload, which are different due to the cross-shard transactions.
Second, the trade-off between workload balance and cross-shard transactions is not addressed, which collaborate to improve scalability.
Third, an adaptive updating mechanism is not introduced for fast execution, which is essential as the algorithm is required to be run periodically.

Instead of graph-based methods, Krol et al. proposed a transaction-level allocation method, Shard Scheduler~\cite{krol2021}. When each transaction comes, the affected accounts are allocated into the least-loaded shard if they satisfy specific criteria based on their historical data.
We follow the stream of graph-based solutions as they capture more global patterns and combinatorially optimize the allocation of a huge amount of accounts.

\section{Problem Formulation}
\label{sec_for}
Following the convention, this work considers that incoming transactions follow similar connection patterns as historical transactions.
We focus on identifying an optimal account-shard mapping using historical transactions.
This section formalizes this problem.
In particular, we first define the blockchain model and the transaction allocation problem, which directly optimizes the system throughput. We model the throughput considering both cross-shard transaction ratio and workload balance, uncovering how these two factors collaborate for better throughput.
Then, we construct a graph to capture the historical transaction patterns and convert the transaction allocation problem to the community detection problem on this transaction graph,  by converting the concepts from the blockchain context into the graph context. 

\subsection{Blockchain Model}
\label{sec_blockchainmodel}
We consider a totally ordered account-based permissionless blockchain as a ledger
$\ledger := \{\Block_{1},\Block_2,...,\Block_n\}$, consisting of a
sequence of $n$ blocks. Each block
$\Block_i := \{\Tx_1, \Tx_2,..., \Tx_{\left|\Block_i\right|}\}$
contains a sequence of transactions, where $|\Block_i|$ is the size of
the $i$-th block. We denote $\setalltx$ and $\setaccount$ as the set
of all transactions and accounts that appeared in $\ledger$, respectively.

A transaction typically contains many data items, such as input/output
accounts, transaction value, gas-related data fields and scripts in
Ethereum. 
As the input and output accounts of a transaction determine the number of involved shards for processing, we define a transaction using only the associated sets $\Inputacc$ and $\outputacc$.  
In particular, we define a transaction as a pair of associated sets,
$\Tx := (\Inputacc,\outputacc)$, where
$\Inputacc, \outputacc \subseteq \setaccount$,
$\Inputacc, \outputacc \neq \varnothing$, and let
$\Transactionacc = \Inputacc \cup \outputacc$.

Incoming transactions are allocated into $k$ shards for concurrent processing.
Each shard is responsible for maintaining the state of a unique set $\shardacc_i$ of accounts, $\shardacc_i \subseteq \setaccount$, $1 < i<k$. 
The accounts $\Transactionacc$ involved in a transaction $\Tx$ may be maintained in multiple shards.
In this case, we consider this transaction is allocated into multiple shards for processing. 
The set of incoming transactions processed in the
  i-th shard is
  $$\transactioninshard_i :=\{\Tx \mid \Transactionacc \cap
  \shardacc_i \neq \varnothing \},$$ which is determined by the
  corresponding $\shardacc_i$. This implies that a transaction will be
  processed by the i-th shard if and only if there exists at least one
  associated account belonging to $\shardacc_i$. Therefore, the
  transaction allocation
  $\{\transactioninshard_1,
  \transactioninshard_2,...,\transactioninshard_k\}$ depends on the
  account allocation. Thus, the goal of this work is to develop an
  effective algorithm to allocate accounts, which guides miners to
  allocate transactions into different shards for better
  scalability. We define it more formally as follows:

  \begin{definition}[Account Allocation]
    Given a blockchain $\ledger$ with $k$ shards and the associated
    parameters $\parameters$, the account allocation algorithm
    $\allocation (\setaccount, \setalltx, \parameters)$ outputs a
    tuple $\{\shardacc_1, \shardacc_2, \ldots, \shardacc_k\}$
    representing the account-shard mapping, such that
 \begin{itemize}
  \item uniqueness: $\forall i,j\in [1,k] \land i \neq j$, $\shardacc_i \cap \shardacc_j = \varnothing$;
  \item completeness, $\setaccount=\cup_{i=1}^k{\shardacc_i}$.
\end{itemize}
\end{definition}

Let $\numrelashards(\Tx)$ be the number of shards involved in
processing a given transaction $\Tx$, defined as follows:
$$\numrelashards(\Tx) := \sum_{i=1}^{k}{\mathds{1}_{\Transactionacc \cap \shardacc_i \neq \varnothing}},$$
where $\mathds{1}_{\Transactionacc \cap \shardacc_i \neq \varnothing}$
is the indicator function, which outputs 1 if
$\Transactionacc \cap \shardacc_i \neq \varnothing$ and 0 otherwise. A
transaction $\Tx$ is cross-shard if $\numrelashards(\Tx)>1$;
otherwise, it is intra-shard.

We utilize a parameter $\Difficulty$ and 1 to represent the workload for a shard to process a cross-shard transaction and an intra-shard transaction, respectively. 
As processing a cross-shard transaction is more complex, $\Difficulty > 1$ generally. 
Note that additional fine-tuning can be applied. For example, the processing workload may differ for input shards and output shards, and for transactions with a different number of affected accounts $\left|\Transactionacc\right|$. 
As exact parameters are application-specific, we only set one parameter $\Difficulty$ for simplicity. This can be easily extended by leveraging different workload parameters based on the specific applications.

As mentioned in Section~\ref{sec_background}, miners are randomly allocated and reshuffled periodically to prevent single-shard take-over attacks. This implies that miners and associated computing resources are uniformly distributed into different shards.
Thus, this work assumes that each shard has an equivalent processing capacity, denoted as a parameter $\lambda$. The capacity parameter $\lambda$ represents the total workload a shard can process. Note that, if all transactions are intra-shard, $\lambda$ equals the number of these transactions a shard can process, as the processing workload for every intra-shard transaction is 1.

\label{sec_definition}

\subsection{The Objective of Account Allocation}
 Given the historical data $\ledger$ as input, the eventual objective of $\allocation$ is to find an account-shard mapping to improve the system throughput. 
 For an account-shard mapping, we introduce the related performance metrics as follows, including cross-shard transaction ratio $\crossratio$, workload balance metrics $\balancemetrics$, system throughput $\throughput$, and transaction confirmation latency $\latency$.

Let  $\setcross$ be the set of all cross-shard transactions in $\ledger$
$$\setcross := \{\Tx \mid \numrelashards(\Tx)>1, \Tx \in \setalltx \}.$$ 
The \textbf{cross-shard transaction ratio} is $\crossratio := \frac{\left|\setcross\right|}{\left|\setalltx\right|}.$

Let $\setintra_i$ and $\setcross_i$ be the set of intra-shard and cross-shard transactions processed in the i-th shard, respectively. 
$$\setintra_i:= \{\Tx \mid \numrelashards(\Tx)=1, \Tx \in \setalltx_i \};$$ 
$$\setcross_i:= \{\Tx \mid \numrelashards(\Tx)>1, \Tx \in \setalltx_i \}.$$ 
Given $\Difficulty$ as a hyper-parameter, the \textbf{workload} of the i-th shard is the total workload to process transactions in it, $$\workload_i= \left|\setintra_i\right|+ \left|\setcross_i\right|*\Difficulty.$$ 
We use the standard deviation to measure the workload balance
\begin{equation}
\label{equ_balance_metric}
    \balancemetrics= \big(\frac{\sum_{i=1}^{k} (\workload_i-\workloadaverage)^{2}}{k} \big)^{0.5},
\end{equation}
where $\workloadaverage=(\sum_{i=1}^{k} \workload_i)/k$ is the mean value of $\workload_i$.

We model the throughput of a blockchain as a mathematical simulation, depending on the historical data $\ledger$ and the specific account-shard mapping.
The overall system \textbf{throughput} is computed as the sum of the throughput of each shard,
\begin{equation}
\label{overallthroughput}
\throughput = \sum_{i=1}^{k} {\throughput_i}.    
\end{equation}
For each shard, there are two cases, namely whether the processing capacity is sufficient to process all allocated transactions or not, i.e., $\workload_i \leq \capacity$ or $\workload_i > \capacity$. 
If the capacity is sufficient, i.e., $\workload_i \leq \capacity$, all transactions in this shard can be processed successfully. 
An intra-shard transaction is considered as a full transaction when counting the throughput of this shard.
An intrinsic difference is that the full processing of a cross-shard transaction $\Tx \in \setcross_i$ requires the cooperation of the other $(\numrelashards(\Tx)-1)$ related shards. Thus, each related shard only counts a share $1/\numrelashards(\Tx)$ of the transaction $\Tx$ in the throughput $\throughput_i$ to prevent redundant counting.
Therefore, when the capacity is sufficient, the throughput of the i-th shard is
$$ \hat{\throughput}_i :=  \sum_{\Tx \in \transactioninshard_i}{\frac{1}{\numrelashards(\Tx)}}.$$

Otherwise, if $\workload_i > \capacity$, we consider that every shard processes transactions chronologically. By this rule, a shard cannot improve its throughput by prioritizing the processing of easy-computing intra-shard transactions.
Therefore, only the proportion $\capacity/\workload_i$ of the total transactions can be successfully processed. 
For the summary of these two cases, the throughput of each shard considering $\capacity$ is 
\begin{equation}
\label{equ_throughput}
   \throughput_i = \begin{cases} \hat{\throughput}_{i}, & \workload_i \leq \capacity \\
\frac{\capacity}{\workload_i}* \hat{\throughput}_{i}, & \workload_i > \capacity \end{cases}.  
\end{equation}

The confirmation \textbf{latency} of each shard is computed by taking the average latency for the transactions in it.
In particular, for a given workload $\workload_i$, the total number of time units required to complete the processing of all transactions is $T = \lceil \workload_i /\capacity \rceil$, where  $\capacity$ is the processing capacity and $\lceil . \rceil$ is the ceiling function.
When $\workload_i <= \capacity$ and $T = 1$, the average latency of all transactions is one time unit, which is equal to $T$.
When $ \workload_i > \capacity$ and $T >= 2$, the average latency differs from $T$.
Only a fraction $(\capacity /\workload_i)$ of transactions are successfully processed in one time unit and another fraction $(\capacity /\workload_i)$ of transactions have the latency of two time units and finally a fraction of $(\workload_i-(T-1)*\capacity)/\workload_i$ of transactions have the latency of $T$. 
To compute the average latency of transactions in the i-th shard, we utilize the integral in terms of the normalized workload $\hat{\workload}_i=\workload_i/\capacity$:
\begin{equation}
\begin{split} 
\label{equ_latency}
\latency_i :&=  \frac{\int_{0}^{\hat{\workload}_i} \lceil x \rceil \,dx}{\hat{\workload}_i}\\
&=\frac{\lfloor  \hat{\workload}_i \rfloor * \lceil \hat{\workload}_i \rceil }{2* \hat{\workload}_i} + \frac{( \hat{\workload}_i-\lfloor  \hat{\workload}_i\rfloor)* \lceil  \hat{\workload}_i\rceil}{ \hat{\workload}_i}
, 
\end{split}
\end{equation}
where $\lfloor . \rfloor$ and $\lceil . \rceil$ are floor and ceiling functions, respectively.
The function $\lceil x \rceil$ indicates the latency for a fraction of transactions, varying from 1 to $T$.
The average latency of all shards is $\latency = (\sum_{i=1}^{k} {\latency_i})/k$.

In conclusion, throughput $\throughput$ and latency $\latency$ reflect the scalability of blockchains. Throughput $\throughput$ is the optimization objective of allocation algorithm $\allocation$. To improve throughput  $\throughput$, the workload of each shard $\workload_i$ is preferred to be lower and balanced with the constraint of $\capacity$, which simultaneously leads to better $\crossratio$ and $\balancemetrics$.

\subsection{Converting to Community Detection}
To solve the transaction allocation problem, we convert it to the community detection problem on graph-structured data. 
In particular, we construct a transaction graph to capture the transaction patterns in $\ledger$ and then convert the blockchain-level performance metrics to this graph.
By this way, we can directly optimize the throughput $\throughput$ to obtain the community structure on this transaction graph. 

The transaction graph models transactions as edges and accounts as nodes.
As the direction of a transaction $\Tx$ does not change $\Transactionacc$ and $\numrelashards(\Tx)$, it can not influence whether $\Tx$ is cross-shard or not.
Thus, this work considers transactions to be undirected for simplicity. 
So, a transaction $\Tx =(\Inputacc,\outputacc)$ can be represented by $\Tx = (\Transactionacc)$.

A transaction $\Tx$ may connect more than two accounts if it is multi-input or multi-output, i.e., $\left|\Transactionacc \right| >2$.
However, in the conventional graph theory, one edge can only connect two nodes. We cope with this problem by separating such a transaction to multiple one-to-one edges with corresponding weights.
The number of corresponding one-to-one edges is
$$\pi(\Tx)=\binom{\left|\Transactionacc\right|}{2}=\frac{\left|\Transactionacc\right|!}{2!*(\left|\Transactionacc\right|-2)!},$$
which is a combination operation to select two accounts from $\left|\Transactionacc\right|$.
Thus, a transaction $\Tx$ assigns the weight of $1/\pi(\Tx)$ to each associated one-to-one edge to ensure the total weights it assigns is one. 
The final weight on each edge is the sum of weights assigned by associated transactions, as detailed in Definition~\ref{def_graph}.

\begin{definition}[Transaction Graph]
\label{def_graph}
The transaction graph $\graph$ of a blockchain $\ledger$ is an undirected weighted graph representing the transaction pattern in $\ledger$. Let $\graph =\left(V, E, W\right)$, where $\setnode = \setaccount$ is the set of nodes, $E \subseteq V \times V$ is the set of edges and $W$ is the set of weights on the edges. 
Each node $v \in V$ in the graph represents an account in $\ledger$; each edge $\{ v,u \}= \{ u,v\} \in E$ represents the existence of a transaction that involves both accounts $v$ and $u$ ; the corresponding weight is the sum of weights assigned by associated transactions, $w_{\{ v,u \}}=w_{\{ u,v\}}= \sum_{\Tx \in \setedgetx_{\{ v,u \}}}{(1/\pi(\Tx))}$, where  $\setedgetx_{\{ v,u\}}= \{\Tx \big| \{v, u\} \subseteq \Transactionacc\}$ is the set of transactions involving both $v$ and $u$. 
\end{definition}

Given transaction graph $\graph$, we aim to derive an account allocation $\allocation$ for better $\throughput$. 
We convert this problem to a community detection problem on $\graph$ and convert the previous blockchain-level performance metrics to the graph-level concepts for optimization.

Community detection is a technique to explore the graph structure and identify communities in the graph \cite{communitydetection}. A community is a set of nodes that should be densely intra-connected whereas sparsely inter-connected. 
 We model $\setaccount_i$ in the i-th shard as a community $V_i$ on $\graph$. 
 Thus, the cross-shard transaction ratio can be written as the inter-community connection ratio,  
 $$\crossratio = \frac{\sum_{v\in V_i, u \in V_j, V_i \neq V_j}{w_{\{ v,u \}}}}{\sum_{v \in V, u \in V}{w_{\{ v,u \}}}}.$$ 
 
The workload of each community $V_i$ can be written as 
\begin{equation}
\label{equ_workloadinshard}
    \workload_i= \sum_{v \in V_i, u \in V_i}\\{w_{\{ v,u \}}+\Difficulty* \sum_{v \in V_i, u \notin V_i}{w_{\{ v,u \}}}}.
\end{equation}
We measure the workload balance by the standard deviation of all $\workload_i$, similar to Equation (\ref{equ_balance_metric}).

We define the throughput of each community $V_i$ on $\graph$. When processing capacity is sufficient for the workload, i.e., $\workload_i<\capacity$, $$\hat{\throughput}_i= \sum_{v \in V_i, u \in V_i}{w_{\{ v,u\}}}+ \frac{\sum_{v \in V_i, u \notin V_i}\\{w_{\{ v,u\}}}}{2}.$$ 
As all the edges in $\graph$ connect to two nodes, the second term indicates that each associated community is responsible for half of the inter-community connection. 
Similarly,  throughput $\throughput_i$ considering capacity and overall throughput $\throughput$ are the same as Equation \ref{equ_throughput} and Equation \ref{overallthroughput}, respectively.
 The latency of each shard $\latency_i$ is the same as Equation \ref{equ_latency}, depending on corresponding $\workload_i$ and $\capacity$. 
 By converting these performance metrics on $G$, an allocation $\allocation$ can be obtained by developing a community detection algorithm to optimize $\throughput$ on $G$.

In conclusion, this section formalizes the transaction allocation problem in sharded blockchains.
We define the blockchain-level performance metrics, including $\throughput$, $\latency$, $\crossratio$ and $\rho$, and convert them to graph-level concepts.
In this way, the transaction allocation problem is converted to the community detection problem to optimize throughput $\throughput$ on graph $G$.

\section{Design challenges}\label{sec_design}
This section discusses the challenges when building a practical transaction allocation method.

\subsection{Deterministic Algorithm}
In a sharded blockchain, miners in each
shard are responsible for processing a fraction of transactions. This
requires different shards to reach a consensus on the set of
transactions allocated to each of them~\cite{han2021security}.
When there are multiple proposals on allocating transactions into
different shards, all miners in the system will need to agree on a
single proposal through a consensus scheme.

Reaching a consensus in
the presence of Byzantine adversaries requires at least 6
communication steps with overall $O(N)$ communication complexity in
streamlined consensus protocols, e.g., HotStuff~\cite{yin2019hotstuff}, or 3 steps with overall $O(N^2)$
communication complexity in classic BFT
protocols, e.g., PBFT~\cite{castro1999practical}~\cite{DecouchantKRY22}. To avoid such a high cost in the
process, the account and transaction allocation algorithm has to be deterministic, resulting in a unique transaction allocation strategy.
This enables different miners in different shards to obtain
the same output for transaction allocation without interacting with
each other, eliminating the need to execute an additional round of
 a consensus algorithm.

\subsection{Adaptive Algorithm}
An adaptive transaction allocation algorithm enables one to update the
allocation result according to the newly accepted transactions, rather
than running the algorithm with all previous transactions as
input. This is particularly important for blockchains of considerable
size. The size of blockchains, such as Bitcoin and Ethereum, is
ever-increasing. At the time of writing, the data size of archive nodes~\cite{chainarchive } and full nodes~\cite{chaindefault} in the Ethereum blockchain is
over 11 TB and 775 GB, respectively, and it has increased by 275 GB and 50 GB per month in the last five months on average.

Each time when new blocks of transactions are recorded on the
blockchain, the transaction allocation should be updated to reflect
the changes in the transaction relations.
If each execution of the transaction allocation algorithm requires all
previous transactions as input, i.e., by simply repeating the
algorithm with a complete set of transactions, it is expensive
both in terms of the required computing power and execution time. This
prevents miners from updating the allocation frequently,
potentially resulting in reduced throughput.

\subsection{Fast Execution}
Each time a new block of transactions is appended to the blockchain,
the transaction graph and the relation between accounts are
updated. To better estimate the community structure from the current updated
graph , the shard allocation algorithm should be executed frequently. The run time of the allocation algorithm determines the upper bound of the frequency to update a new account-shard mapping.

Let $t_b$ be the block generation time interval, and  $t_r$ be the execution time for the allocation algorithm.
 $t_r$ determines the latency of applying the newly
derived account-shard mapping. If the allocation algorithm is executed
according to the $i$-th block, then the output mapping can either be
applied after the block $i+\ceil{\frac{t_r}{t_b}}$ without terminating
the block generation process, or be applied to the $(i+1)$-th block
by waiting for $t_r$ time before processing the next block. The former
introduces latency in applying the latest result to the block
generation, and the latter introduces latency in generating the next
block. Thus, an expected $t_r$ should be less than $t_b$ to enable the updating frequency of account-shard mapping to reach the upper bound of every block.

 \section{Methodology}\label{sec_method}
This section introduces \NAME{}, a transaction allocation scheme to dynamically identify an account-shard mapping by optimizing the throughput. We provide a discussion on the integration to existing protocols in Section~\ref{sec_integration}.

 \subsection{Model Overview}
 \NAME{} has two main components: the global algorithm \globalNAME{} and the adaptive algorithm \adaNAME. 
 The former optimizes the throughput based on all historical transaction data.  
The latter inherits from the previous allocation results and obtains the approximate allocation only based on the newly-committed transaction data for faster execution. 
   \adaNAME{} and \globalNAME{} are executed periodically with different frequencies $\updateada$ and $\updateglo$, respectively. 
   \adaNAME{} is executed more frequently than \globalNAME{} for fast updating, i.e., $\updateada < \updateglo$. 
   The lower bounds of $\updateada$ and $\updateglo$ are determined by the running time of \adaNAME{} and \globalNAME, respectively.
Theoretically, as an approximate algorithm, \adaNAME{} introduces the performance loss. Periodic execution of \globalNAME{} is required to bound this performance loss.

  The hyperparameters of \NAME{} are updating frequencies $\updateada$ and $\updateglo$, number $\numshard$ of shards, convergence threshold $\convergence$, workload $\Difficulty$ of processing a cross-shard transaction and processing capacity $\capacity$ of each shard.

 \subsection{\globalNAME}
 \globalNAME{} algorithm consists of two phases, namely initialization and optimization. 
In the initialization phase, we leverage a classic community detection algorithm, the Louvain method, to obtain the initial account-shard mapping.
In the optimization phase, nodes are re-distributed to form the final account-shard mapping by optimizing the throughput $\throughput$.
We will detail \globalNAME{} in Algorithm~\ref{algorithm_global}, after discussing the optimization method as follows.

 \textbf{Optimization Method.}  \globalNAME{} sequentially loops for every node for optimization. 
A node $v$ firstly leaves its previous community and then joins the community with the largest throughput gain, which is computed by Equation~\ref{throughput_gain}. We introduce the computation in detail as follows.

When a node $v$ leaves community $V_p$ and joins community $V_q$, only two communities change their throughput, $V_p$ and $V_q$, which will be proved in Lemma~\ref{hypothesis}. 
The throughput gain of $V_q$ during the joining procedure is computed by:
\begin{equation}
\label{throughput_join}
    \Delta_{i,join} \throughput_q = \throughput'_{q} - \throughput_{q},
\end{equation}
where $\throughput'_{q}$ is the throughput of $V_q$ after $v$ joining it.

Similar to Equation~(\ref{equ_throughput}),  $\throughput'_{q}$ is computed by:
\begin{equation}
\label{equ_throughputafter}
    \throughput'_{q} = \begin{cases} \hat{\throughput}'_{q} , &  \workload'_{q} \leq \capacity \\
\frac{\capacity}{ \workload'_{q}}* \hat{\throughput}'_{q}
, &  \workload'_{q}> \capacity \end{cases},
\end{equation}
where $ \workload'_q$ is the workload in $V_q$ after moving and $ \hat{\throughput}'_{q}$ is the throughput when capacity is sufficient after moving.

Self-loop transactions may happen in Blockchains. Taking Ethereum as an example, when an account wants to withdraw a pending transaction, one method to achieve this is to send a transaction to itself with a high gas price.
Thus, there are three kinds of nodes connected to $v$: self-loop, nodes in $V_q$ and nodes in other communities. Let the weight between node $v$ and community $V_q$ be the sum of weight between node $v$ and all nodes in  $V_q$, 
$$ w_{\{ v,V_q\}} = \sum_{u \in V_q}{w_{\{ v, u\}} }.$$
The workload in $V_q$ after moving is computed by:
\begin{align*} 
 \workload'_q &=  \workload_q + (1-0)*w_{\{ v,v\}} \\
&+(\Difficulty-0) * (w_{\{ v,V/V_q\}}-w_{\{ v,v\}})\\
&+(1-\Difficulty)*w_{\{ v,V_q\}},
\end{align*}
where $w_{\{ v,v\}}$ is the self-loop of node $v$. 
From the perspective of the q-th shard, the first term indicates that the self-loop connections of $v$ are turned to intra-shard workload.
The second term indicates that the connections between $v$ and other communities are turned to cross-shard workload. 
The third term indicates that the connections between $v$ and $V_q$ are turned from cross-shard to intra-shard in the q-th shard.

Similarly, when processing capacity $\capacity$ is sufficient, the throughput in $V_q$ after moving is computed by
\begin{align*} 
 \hat{\throughput}'_{q} &=  \hat{\throughput}_{q}+w_{\{ v,v\}}+\frac{w_{\{ v,w_{\{ v,V/V_q\}}-w_{\{ v,v\}}\}}}{2} +\frac{w_{\{ v,V_q\}}}{2}\\
                               &= \hat{\throughput}_{q}+ w_{\{ v,v\}}+\frac{w_{\{ v,V/v\}}}{2}.
\end{align*}

The above discussions provide a method to computed $\Delta_{i,join} \throughput_q $ by Equation~(\ref{throughput_join}). 
It is similar to compute the throughput gain $\Delta_{i,leave} \throughput_p$ of $V_p$ in the leaving procedure. 
Note that the difference is that $v$ belongs to $V_p$, i.e., $v \in V_p$. 
Specifically, 
$$ \workload'_p = \workload_p-w_{\{ v,v\}} 
-\Difficulty * (w_{\{ v,V/V_p\}})
+(\Difficulty-1)*w_{\{ v,V_p/v\}}.$$

$$ \hat{\throughput}'_{p} = \hat{\throughput}_{p} -w_{\{ v,v\}}-\frac{w_{\{ v,V/v\}}}{2}.$$

\begin{lemma}[]
\label{hypothesis}
 When node $v$ moves from community $V_p$ to community $V_q$, throughput $\throughput_j$ of any other community $V_j$ remains the same for $\forall j \in [1,k] \land  j \neq p \land  j \neq q$.
\end{lemma}

 \textbf{Proof:} For $j \neq p$ and $j \neq q$, $\throughput'_j$ can be similarly computed by Equation~(\ref{equ_throughputafter}). Because $v$ does not leave and join $V_j$ in this move, we have 
\begin{align*} 
\Delta \workload_j &= (0-0)*w_{\{ v,v\}} \\
&+(0-0) * (w_{\{ v,V/V_j\}}-w_{\{ v,v\}})\\
&+(\Difficulty-\Difficulty)*w_{\{ v,V_j\}}\\
&=0,
\end{align*}
and similarly, $\Delta \hat{\throughput}_{j} = 0$. Thus, $\throughput'_j=\throughput_j$  by Equation~(\ref{equ_throughputafter}).

Therefore, the overall throughput gain when $v$ moves from $V_p$ to $V_q$,
\begin{equation}
\label{throughput_gain}
    \Delta_{(i,p,q)} \throughput = \Delta_{i,leave} \throughput_p + \Delta_{i,join} \throughput_q.
\end{equation}

To accelerate the algorithm, we identify a small searching space of the optimal community for each node $v$, called candidate communities $\candidate_{v}$ in this work.
Supposing $v \in V_p$ before moving, a straw-man proposal is that a node $v$ can  join any $V_q \in \{V_1, V_2,...,V_k\}/{V_p}$. However, most nodes are not densely-connected, i.e., they tend to only connect to a small number of communities. 
By this observation, we set candidate communities only to include the communities connecting to the node $v$  rather than all the communities.
\begin{equation}
\label{equ_can}
    \candidate_{v} = \{V_j \mid w_{\{ v,V_j\}} \neq 0 ,V_j \in \{V_1, V_2,...,V_k\}/{V_p}\}.
\end{equation}

 The pseudo-code of \globalNAME{} in Algorithm~\ref{algorithm_global} illustrates the detailed procedure.
 The inputs of \globalNAME{} include transaction graph $G$ and associated hyperparameters.

 \begin{algorithm}[tbp]
 \begin{scriptsize}
 \caption{\globalNAME}
 \label{algorithm_global}
 \begin{algorithmic}[1]
\REQUIRE
  $\graph =\left(V, E, W\right)$: transaction graph;  
  $\Difficulty$: workload of processing a cross-shard transaction; 
  $\numshard$: number of shards; 
  $\capacity$: processing capacity of each shard; 
  $\convergence$: convergence threshold. 
  \ENSURE
$\{V_1, V_2,...,V_k\}$: allocation result for $V$.
\newline

\STATE Initialization: $\{V_1, V_2,...,V_k,...,V_l\}$ using the Louvain method. 
\FOR{$v\in V_{small}$}
    \STATE Compute $\candidate_{v}$  using Equation~(\ref{equ_can}).
    \IF{$\candidate_{v} = \varnothing$}
    \STATE $\candidate_{v} = \{V_1, V_2,...,V_k\}$
    \ENDIF
         \STATE Find the community with the largest throughput gain of joining procedure using Equation (\ref{throughput_join}): $\arg \max_{V_q \in \candidate_{v}}{(\Delta_{i,join} \throughput_q)}$.
        \STATE Update allocation: $v \rightarrow V_q$
    \ENDFOR
\WHILE{$ \Delta \throughput  >= \convergence$}
    \STATE $\Delta \throughput  = 0$
    \FOR{$v\in V$}
    \STATE Compute $\candidate_{v}$  using Equation~(\ref{equ_can}).
    \STATE Find the community with the largest throughput gain using Equation (\ref{throughput_gain}): $\arg \max_{V_q \in \candidate_{v}}{(\Delta_{(i,p,q)} \throughput)}$.
    \IF{$\Delta_{(i,p,q)} \throughput > 0$}
    \STATE Update allocation: $v \rightarrow V_q$; $\Delta \throughput += \Delta_{(i,p,q)} \throughput$.
    \ENDIF
    \ENDFOR
    \ENDWHILE
    
\RETURN $\{V_1,V_2,...,V_k\}$

  \end{algorithmic}
  \end{scriptsize}
\end{algorithm}

\textbf{Initialization Phase.}  Lines 1 -- 9 in Algorithm~\ref{algorithm_global} introduce the initialization phase. In particular, we leverage the Louvain method \cite{blondel2008fast} for the initialization in this work. The Louvain method is a hierarchical community detection algorithm that can process large-scale graphs efficiently. 
The Louvain method can derive an initialization of account-shard mapping $\{V_1, V_2,\ldots, V_l\}$. 
However, the number $l$ of communities is automatically identified by the Louvain method instead of being manually preset.
Due to the long-tailed distribution of nodes, the Louvain method usually tends to derive a significant number of communities, i.e., $l > k$.
In the uncommon situation $l < k$, we generate $\{V_1, V_2,\ldots,V_l\, \ldots,V_k\}$ to initialize $k$ shards, where $V_j = \varnothing$ for $j>l$.

If $l > k$, let $\{V_1, V_2,\ldots, V_k,\ldots, V_l\}$ be the results identified by the Louvain method, where $\{V_1, V_2,\ldots,V_k\}$ are the largest $k$ communities w.r.t. workload $\workload_i$.
We denote the set of nodes in other small communities as $V_{small} = \cup_{j=k+1}^{l}{V_j}$.
The initialization phase moves nodes in small communities into one of the largest $k$ communities to derive account-shard mapping with $k$ shards.
In lines 2--9, \globalNAME{} loops for the nodes in small communities $V_{small}$.
The set of candidate communities $\candidate_{v}$ is computed for the node $v$ by Equation~(\ref{equ_can}).
If $v$ does not connect to any large community, \globalNAME{} specifically forces $\candidate_{v} = \{V_1, V_2,...,V_k\}$ to ensure $v$ leaving the previous community.
In line 7, \globalNAME{} searches $V_q \in \candidate_{v}$  with the largest throughput gain using Equation (\ref{throughput_join}).
Note that, in this step, the throughput gain is computed only using $V_q$, as only the largest $k$ communities remain after the initialization phase.
 $V_p \subseteq V_{small}$ will finally be an empty set, which is irrelevant to the throughput $\throughput$. 
Finally, we update the allocation and the associated workload $\workload_q$ and throughput $\throughput_q$.

\textbf{Optimization Phase.}
 After all nodes are moved into the largest $k$ communities,  \globalNAME{} loops for all nodes to optimize the throughput gain by  Equation (\ref{throughput_gain}). 
 Lines 10 -- 20 illustrate the optimization phase in detail.
 Similarly, we first compute $\candidate_{v}$ for each node $v$.
The difference is that we allow $\candidate_{v} = \varnothing$. In this case, we do not move $v$ as it does not connect to any other community except for $V_p$. Then, \globalNAME{} searches the optimal community $V_q \in \candidate_{v}$ using Equation (\ref{throughput_gain}) considering both the leaving procedure of $V_p$ and the joining procedure of $V_q$. Note that node $v$ is allowed to stay in the current community $V_p$ and it is moved to $V_q$ only when $\Delta_{(i,p,q)} \throughput > 0$. This throughput optimization procedure loops until the throughput gain is less than a convergence threshold $\convergence$. Finally, \globalNAME{} obtains the converged result of account-shard mapping within $\numshard$ communities.

 The time complexity for initialization is $O(Nlog(N))$, where $N = \left|V\right|$~\cite{lancichinetti2009community}. The time complexity for optimization is $O(N\numshard)$. Thus, the total time complexity is $O(Nlog(N)+N\numshard)$. To further reduce the time complexity, \adaNAME{} is proposed and introduced in the next section.
 
 Every step of \globalNAME{} is deterministic. Thus, the algorithm is deterministic when given an identical ordered node sequence to loop.
 The Louvain method for initialization is also deterministic with the ordered node sequence.
 The hash value of the accounts can determine the order of node sequence in real-world applications.

\subsection{\adaNAME}
 Algorithm~\ref{algorithm_global} loops for all nodes in historical transactions for precise allocation. However, the scale of all nodes keeps increasing with blocks generating, which limits the running time of \globalNAME. The expected complexity of the allocation algorithm is $O(1)$ w.r.t. $N$, which means the running time is constant when blocks accumulate.

To cope with this problem, we propose an adaptive algorithm \adaNAME{} with faster execution to update account-shard mapping more frequently. 
Instead of using the entire historical transaction data, \adaNAME{} is only based on the previous account-shard mapping and the newly-committed transactions, resulting in lower computational complexity.  

Let $\graph =\left(V, E, W\right)$ be the transaction graph of the latest $\ledger$. We denote $\adaV$ as the set of all nodes that appear in the newly generated $\updateada$ blocks and $\adaV \subseteq V$. Let $\{V_1,V_2,...,V_k\}$ be the allocation of the last update and $\cup_{j=1}^{k}{V_j} \cup \adaV = V$. The pseudo-code of \adaNAME{} is shown in Algorithm \ref{algorithm_ada}.

\begin{algorithm}[tbp]
 \begin{scriptsize}
 \caption{\adaNAME}
 \label{algorithm_ada}
 \begin{algorithmic}[1]
\REQUIRE
  $\graph =\left(V, E, W\right)$: new transaction graph; 
  $\adaV$: nodes appeared in  newly-committed transactions;
  $\{V_1,V_2,...,V_k\}$: previous allocation result;
  $\Difficulty$: workload of processing a cross-shard transaction; 
  $\capacity$: processing capacity of each shard; 
  $\convergence$: convergence threshold. 
  \ENSURE
$\{V_1, V_2,...,V_k\}$: new allocation result for $V$ that $\cup_{j=1}^{k}{V_j} = V$.
\newline

\FOR{$v\in \adaV-(\cup_{j=1}^{k}{V_j})$}
        \STATE Compute $\candidate_{v}$  using Equation~(\ref{equ_can}).
        \IF{$\candidate_{v} = \varnothing$}
        \STATE $\candidate_{v} = \{V_1, V_2,...,V_k\}$
        \ENDIF
        \STATE Find the community with the largest throughput gain of joining procedure using Equation (\ref{throughput_join}): $\arg \max_{V_q \in \candidate_{v}}{(\Delta_{i,join} \throughput_q)}$.
        \STATE Update allocation: $v \rightarrow V_q$
        \ENDFOR
\WHILE{$\Delta \throughput  >= \convergence$}
    \STATE $\Delta \throughput  = 0$
    \FOR{$v\in \adaV$}
    \STATE Find the community with the largest throughput gain using Equation (\ref{throughput_gain}): $\arg \max_{V_q \in \candidate_{v}}{(\Delta_{(i,p,q)} \throughput)}$.
    \IF{$\Delta_{(i,p,q)} \throughput > 0$}
    \STATE Update allocation: $v \rightarrow V_q$; $\Delta \throughput += \Delta_{(i,p,q)} \throughput$.
    
    \ENDIF
    \ENDFOR

\ENDWHILE
\RETURN $\{V_1,V_2,...,V_k\}$
\end{algorithmic}
\end{scriptsize}
\end{algorithm}

\adaNAME{} loops for nodes in $\adaV$ instead of the entire set of nodes $V$.
A node $v \in \adaV$ can either be existed in the previous allocation or not. 
In lines 1 -- 8, \adaNAME{} first allocates new nodes.
When $v$ is new, i.e. $v \not\in \cup_{j=1}^{k}{V_j}$, the condition is similar to the small communities in \globalNAME.
If the new node $v$ connects to existing communities, i.e. $\candidate_{v} \neq \varnothing$,  \adaNAME{} optimizes throughput gain using Equation (\ref{throughput_join}) similarly.
Otherwise, if it does not connect to any existing community, \adaNAME{} sets all communities as candidate communities. 
In lines 9 -- 18, after all new nodes have been allocated, \adaNAME{} loops for all $v \in \adaV$ for optimization.

\adaNAME{} is also deterministic and the time complexity is $O(|\adaV|\numshard)$. As the updating gap $\updateada$ is constant, $|\adaV|$ is quasi-constant. It is much smaller than $|V|$ and does not accumulate when blocks are generated. Thus, the time complexity is constant with the scale of blockchain data, i.e. $O(1)$.

\section{Experimental Results}\label{sec_exp}
We conduct extensive experiments on real-world Ethereum data for simulations.
For both \globalNAME{} and \adaNAME, we evaluate the quality of identified communities and show that the algorithms can achieve a low cross-shard transaction ratio and keep a balanced workload, resulting in throughput improvement. We also show the running time comparison, which achieves improvement in scale.
The experiments are implemented using Python 3.8 on a computing cluster node with Intel Xeon Gold 6150 CPU and 250 GB memory.

\begin{figure}[h]
    \centering
    \includegraphics[width=0.4\textwidth]{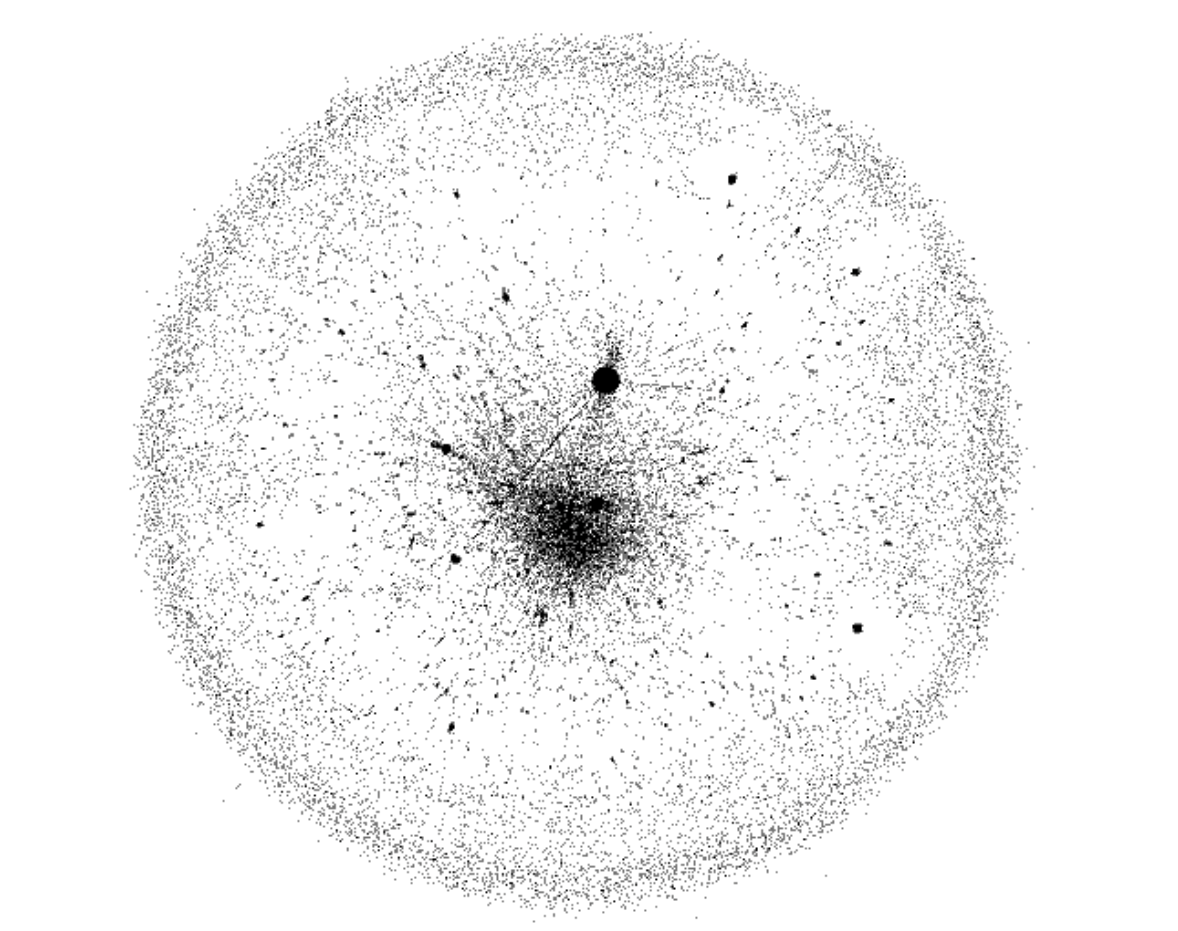}
    \caption{\small Visualization of the graph structure of the Ethereum dataset. The dots represent accounts and the links represent transactions. The scale of a dot represents its activeness. The distance between dots represents the number of associated transactions between them.}
    \label{vis}
\end{figure}

\begin{figure*}[ht]
\captionsetup[subfigure]{justification=centering}
    \centering
    \begin{subfigure}{0.17\textwidth}
     \centering
     \includegraphics[width=1\textwidth]{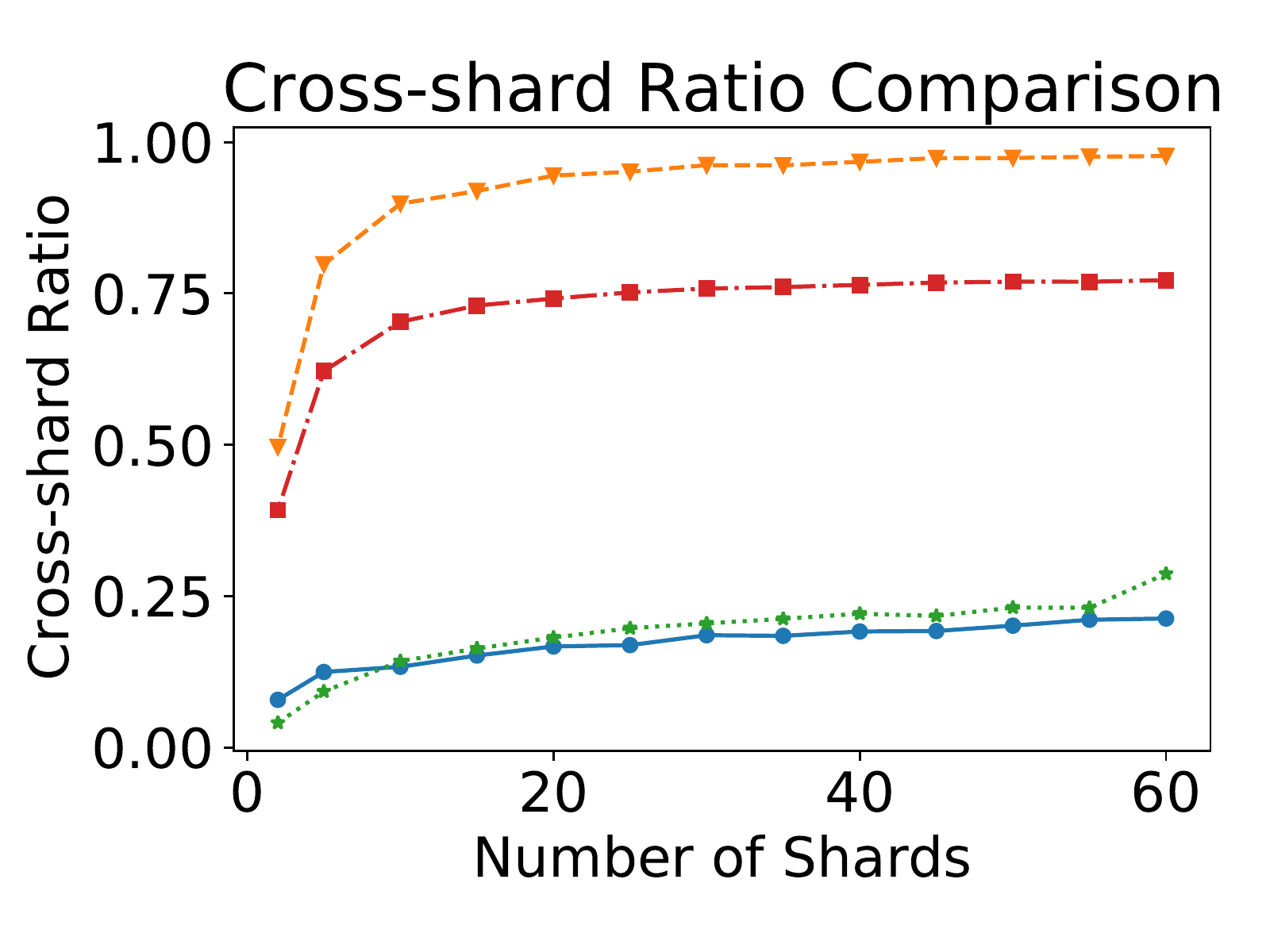}
     \caption{ $\Difficulty = 2$}
    \end{subfigure}
    \begin{subfigure}{0.17\textwidth}
     \centering
      \includegraphics[width=1\textwidth]{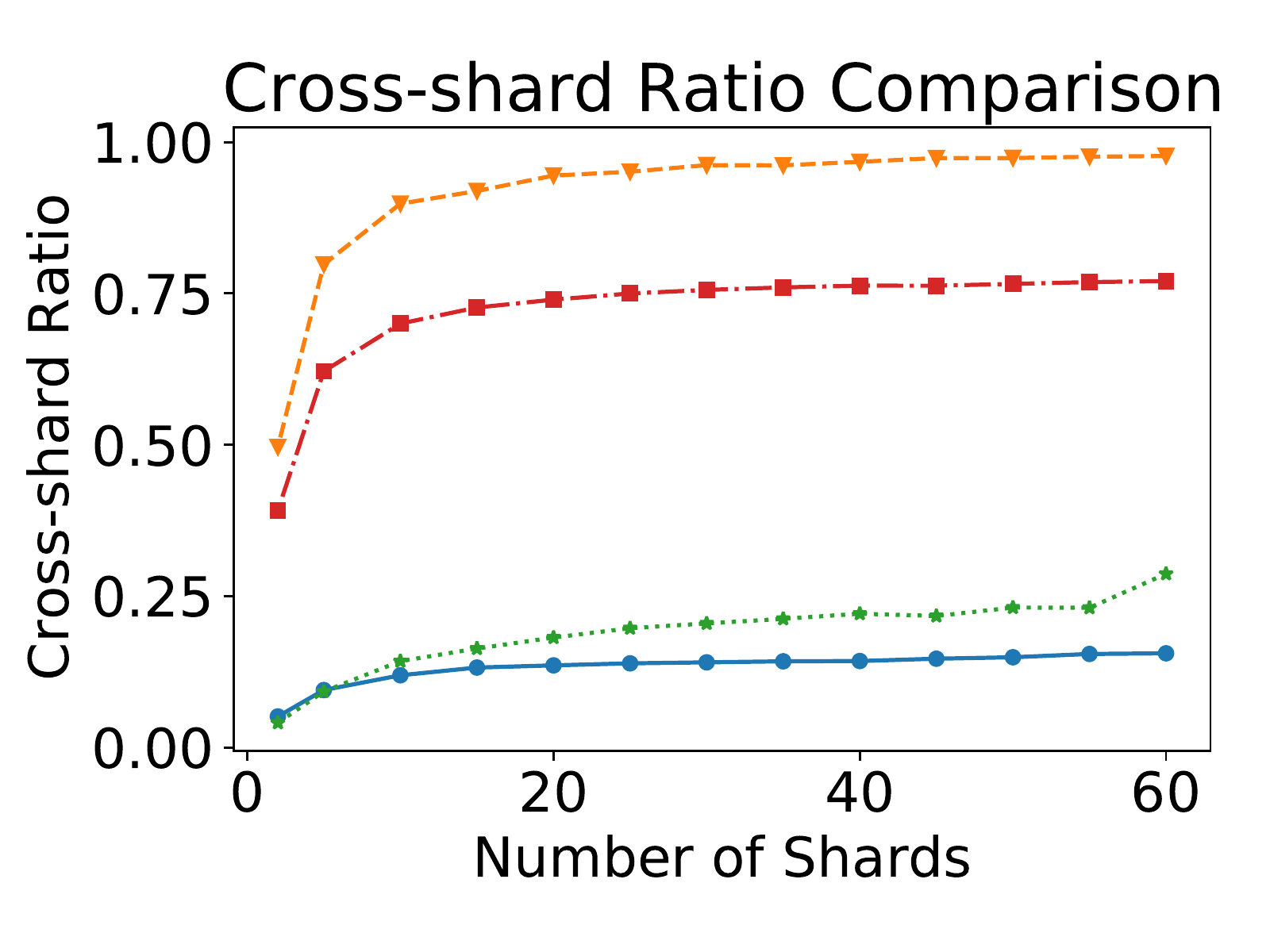}
      \caption{ $\Difficulty = 4$}
    \end{subfigure}
    \centering
    \begin{subfigure}{0.17\textwidth}
     \centering
     \includegraphics[width=1\textwidth]{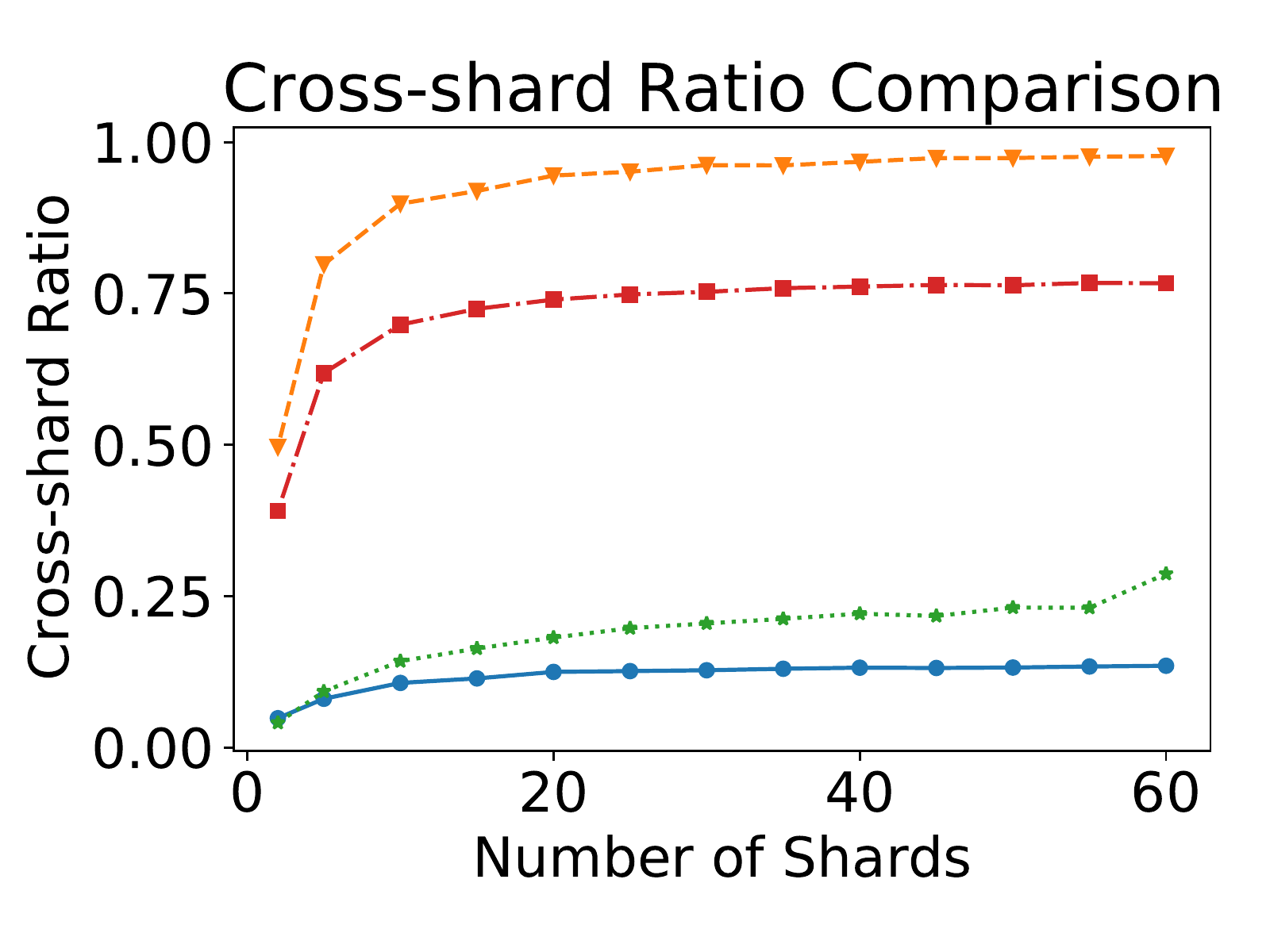}
     \caption{ $\Difficulty = 6$}
    \end{subfigure}
    \begin{subfigure}{0.17\textwidth}
     \centering
      \includegraphics[width=1\textwidth]{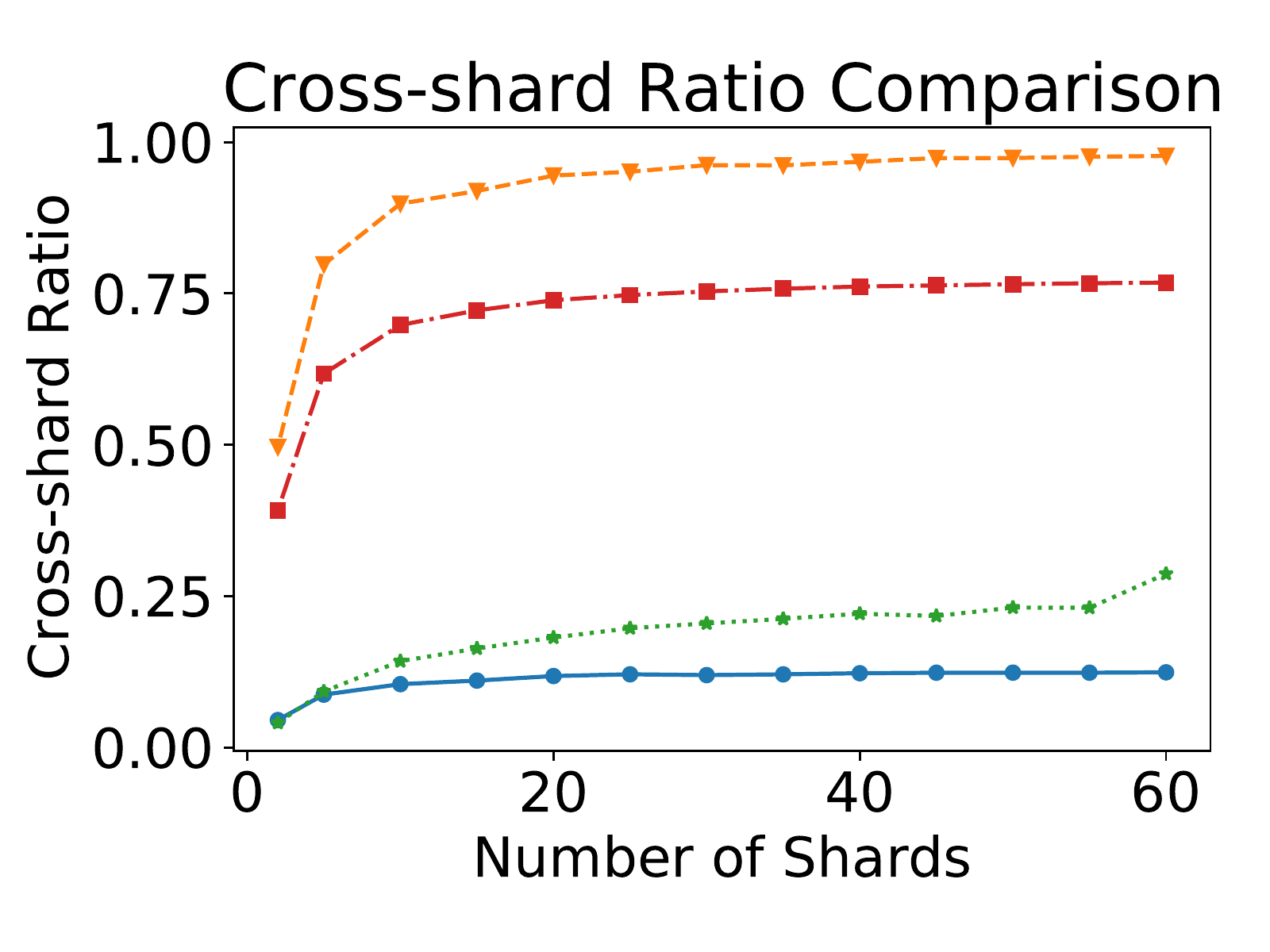}
      \caption{ $\Difficulty = 8$}
    \end{subfigure}
    \begin{subfigure}{0.26\textwidth}
     \centering
      \includegraphics[width=1\textwidth]{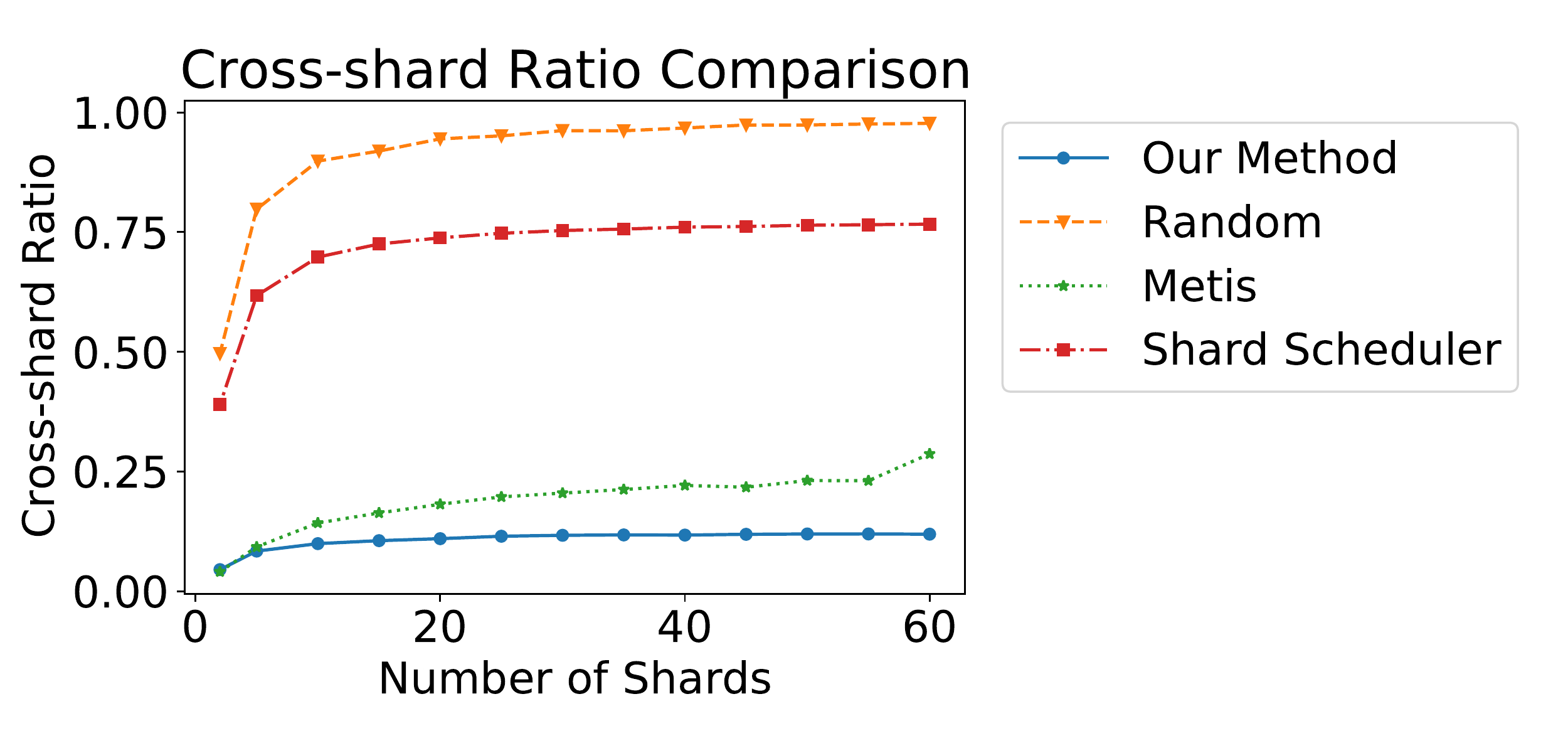}
      \caption{ $\Difficulty = 10$}
    \end{subfigure}
    \caption{Cross-shard transaction ratio comparison with various $\numrelashards$ and $\Difficulty$. }
    \label{global_ratio}
\end{figure*}

\begin{figure*}[ht]
\captionsetup[subfigure]{justification=centering}
    \centering
    \begin{subfigure}{0.17\textwidth}
     \centering
     \includegraphics[width=1\textwidth]{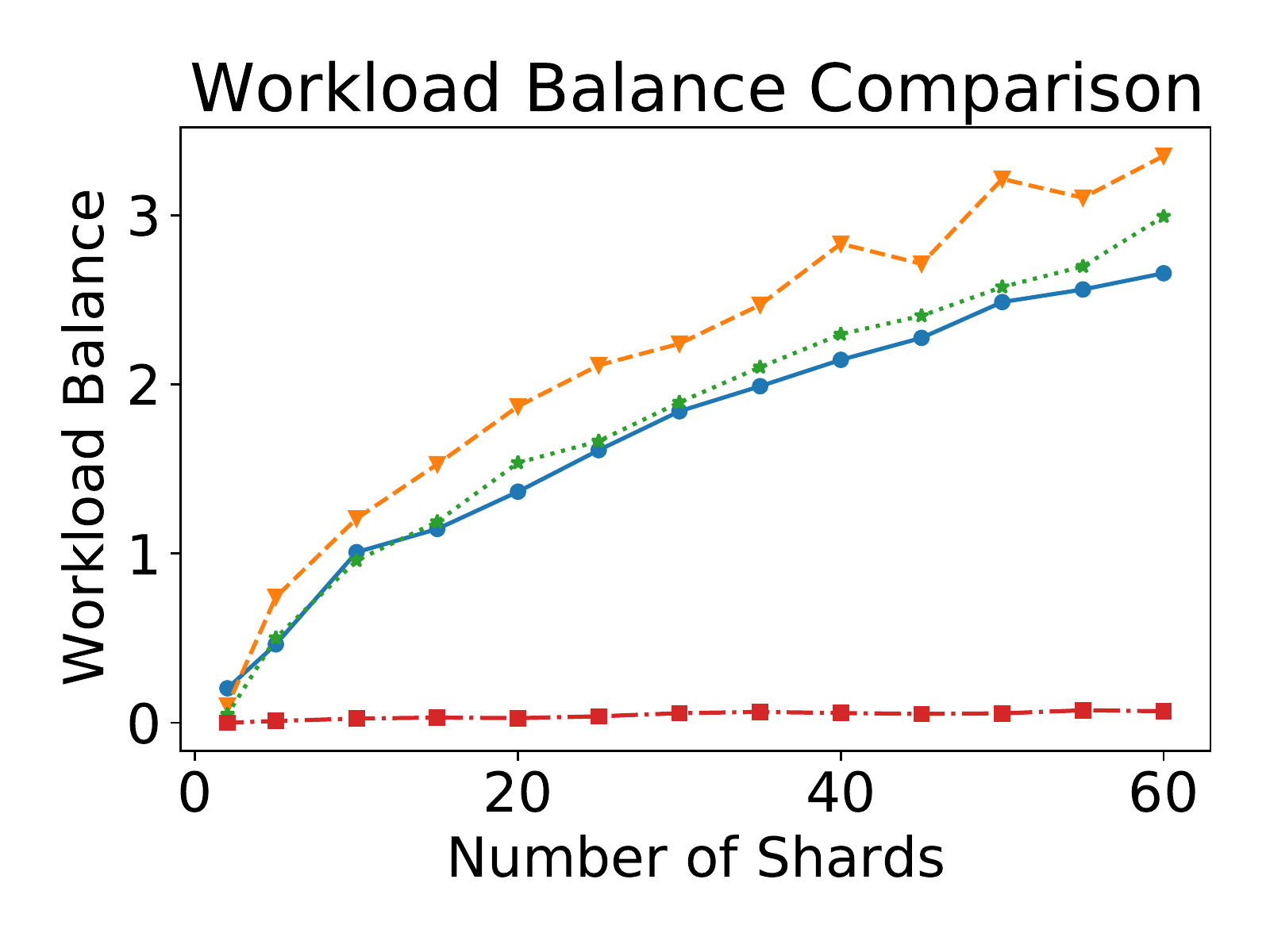}
     \caption{ $\Difficulty = 2$}
     \end{subfigure}
    \begin{subfigure}{0.17\textwidth}
     \centering
      \includegraphics[width=1\textwidth]{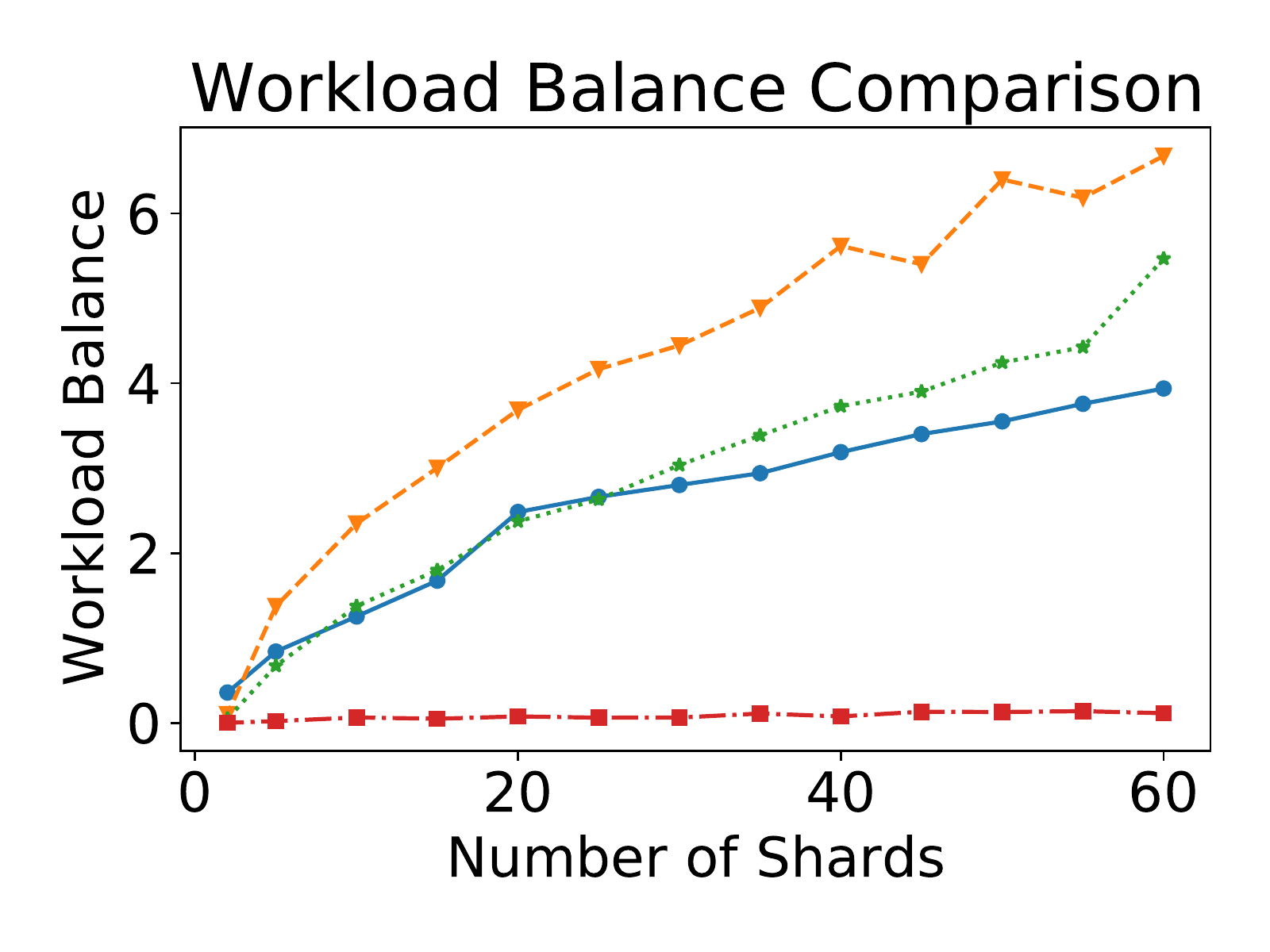}
      \caption{ $\Difficulty = 4$}
    \end{subfigure}
    \centering
    \begin{subfigure}{0.17\textwidth}
     \centering
     \includegraphics[width=1\textwidth]{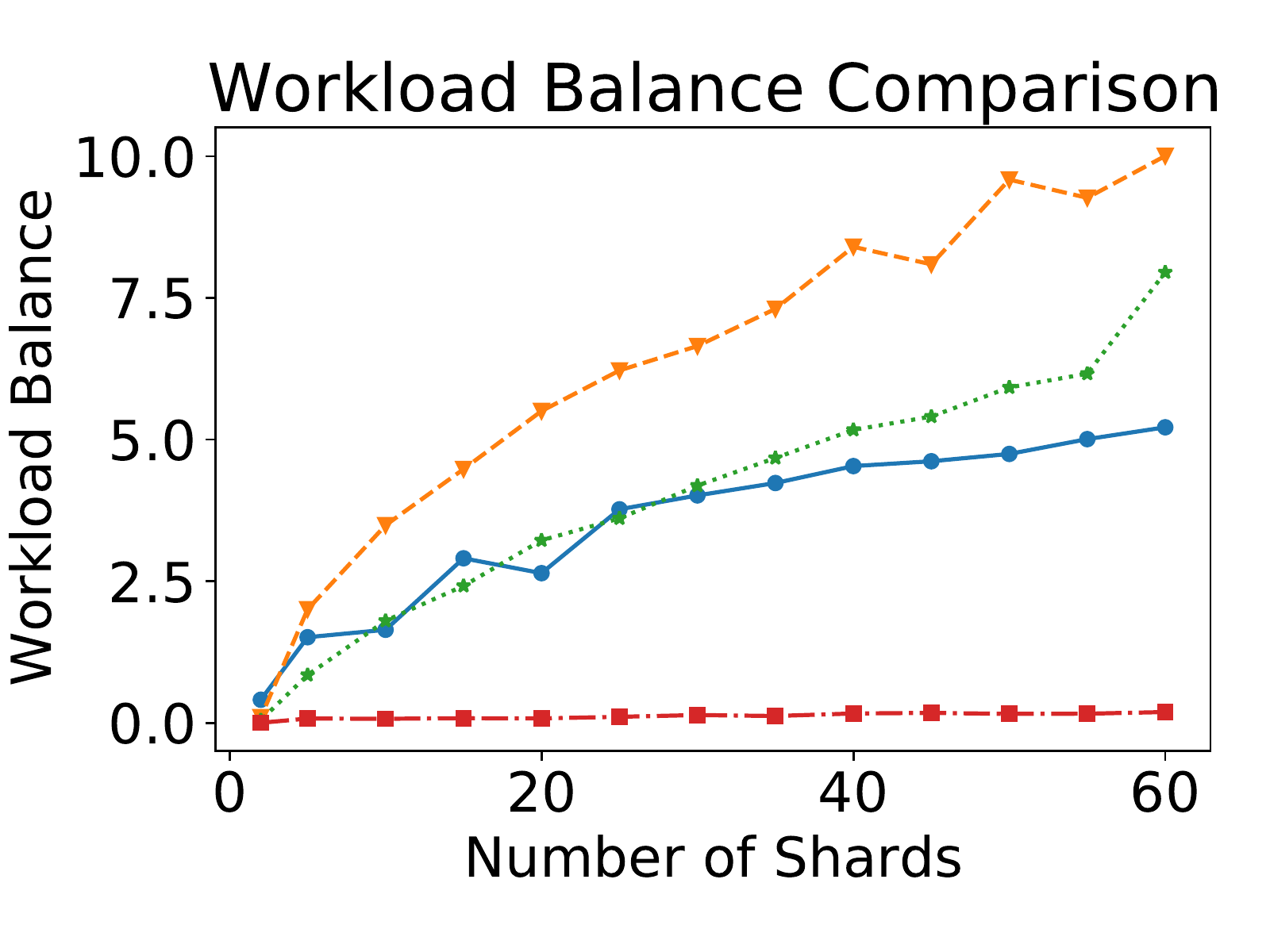}
     \caption{ $\Difficulty = 6$}
    \end{subfigure}
    \begin{subfigure}{0.17\textwidth}
     \centering
      \includegraphics[width=1\textwidth]{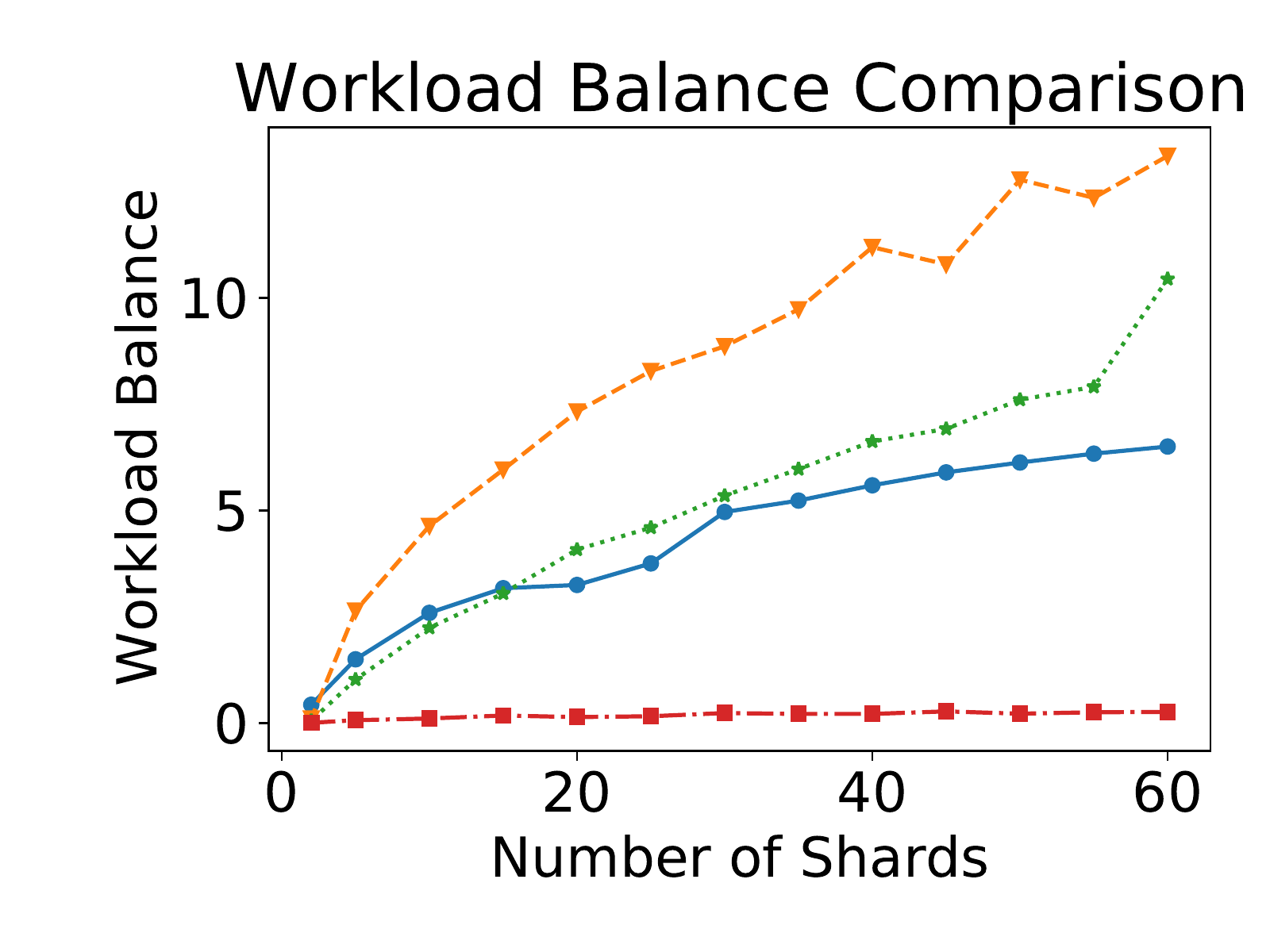}
      \caption{ $\Difficulty = 8$}
    \end{subfigure}
    \begin{subfigure}{0.26\textwidth}
     \centering
      \includegraphics[width=1\textwidth]{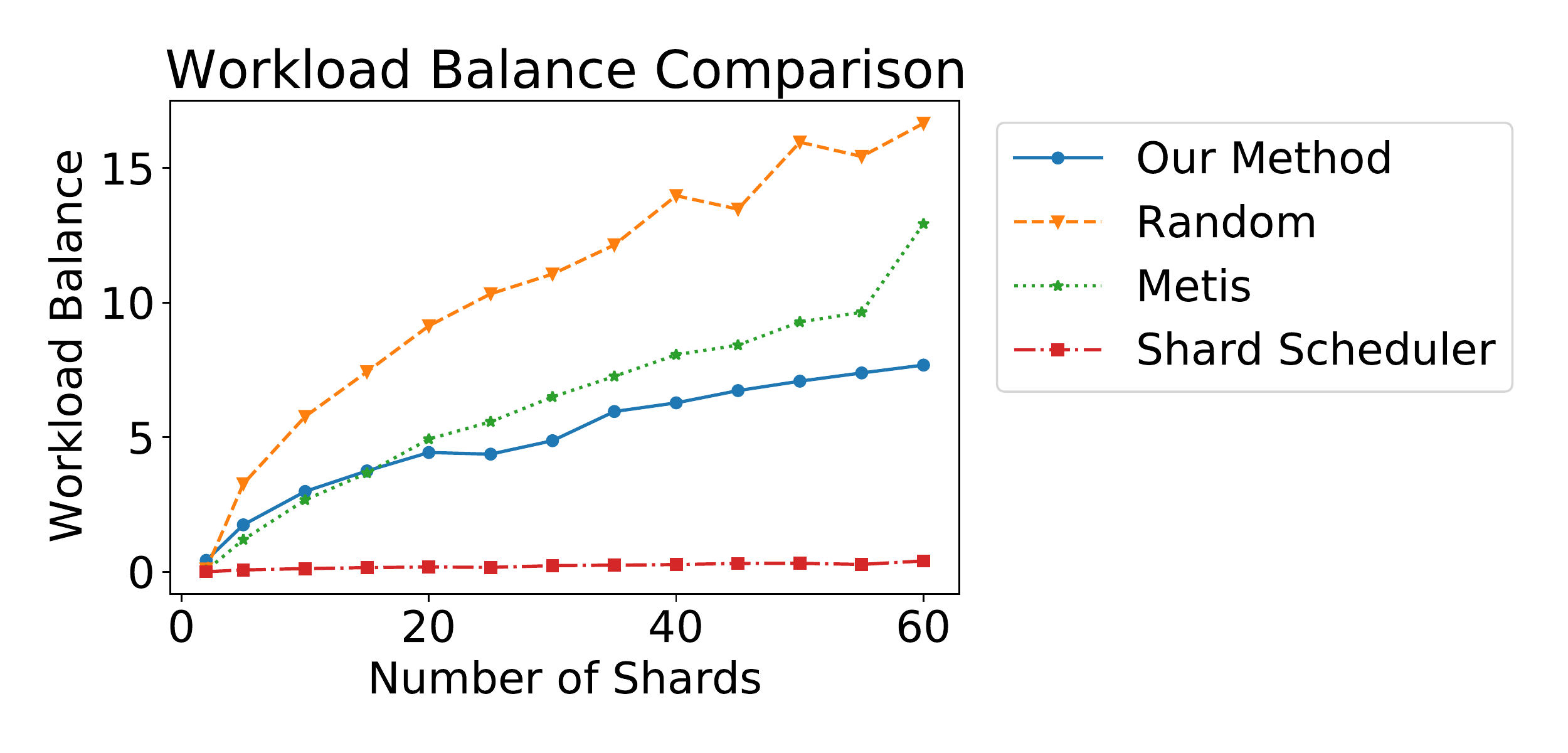}
      \caption{ $\Difficulty = 10$}
    \end{subfigure}
    \caption{Workload balance comparison with various $\numrelashards$ and $\Difficulty$. }
    \label{global_balance}
\end{figure*}

\begin{figure}[ht]
\captionsetup[subfigure]{justification=centering}
    \centering
    \begin{subfigure}{0.24\textwidth}
     \centering
     \includegraphics[width=1\textwidth]{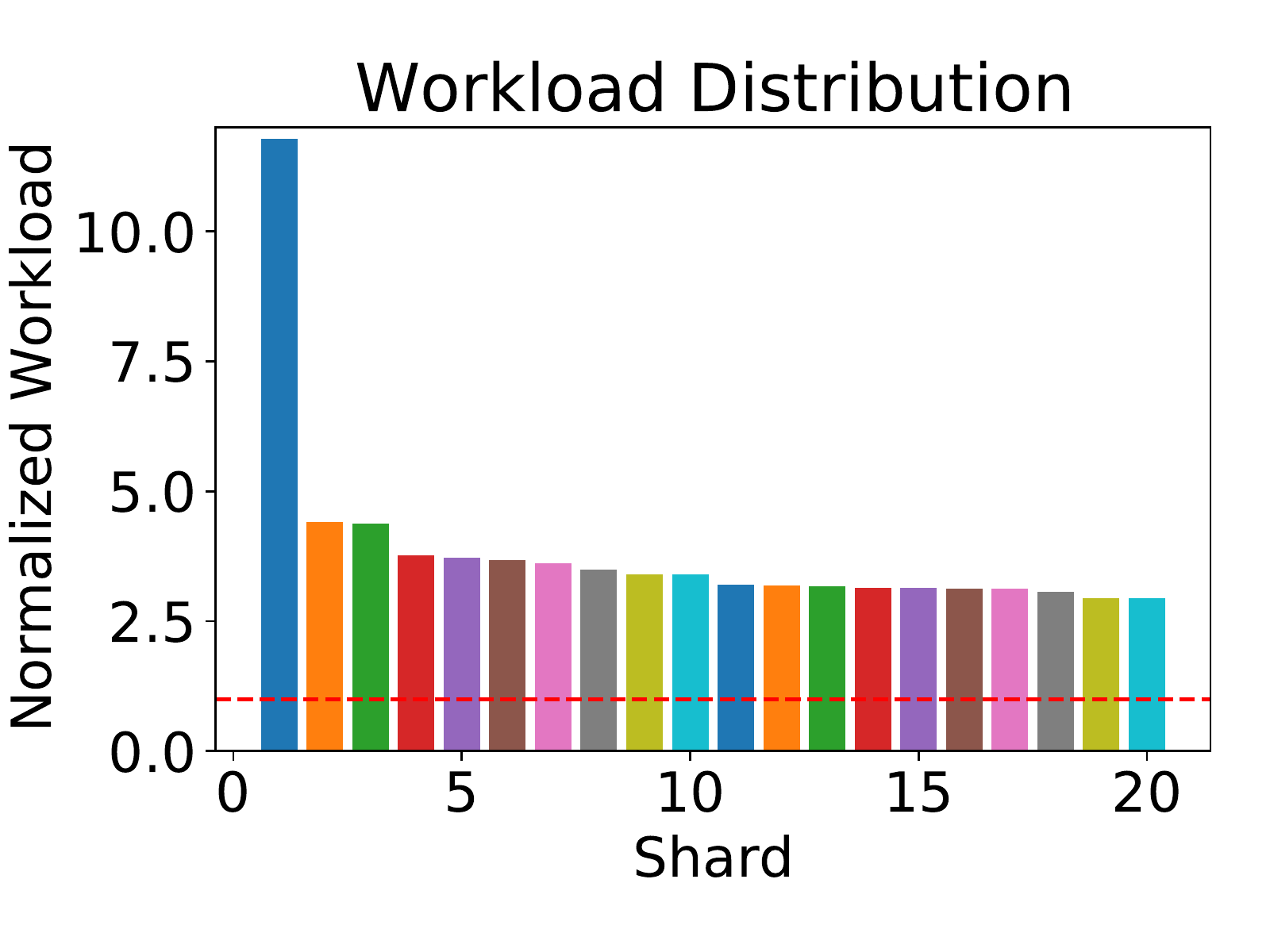}
     \caption{Random}
      \label{Randomdistribution}
     \end{subfigure}
    \begin{subfigure}{0.24\textwidth}
     \centering
      \includegraphics[width=1\textwidth]{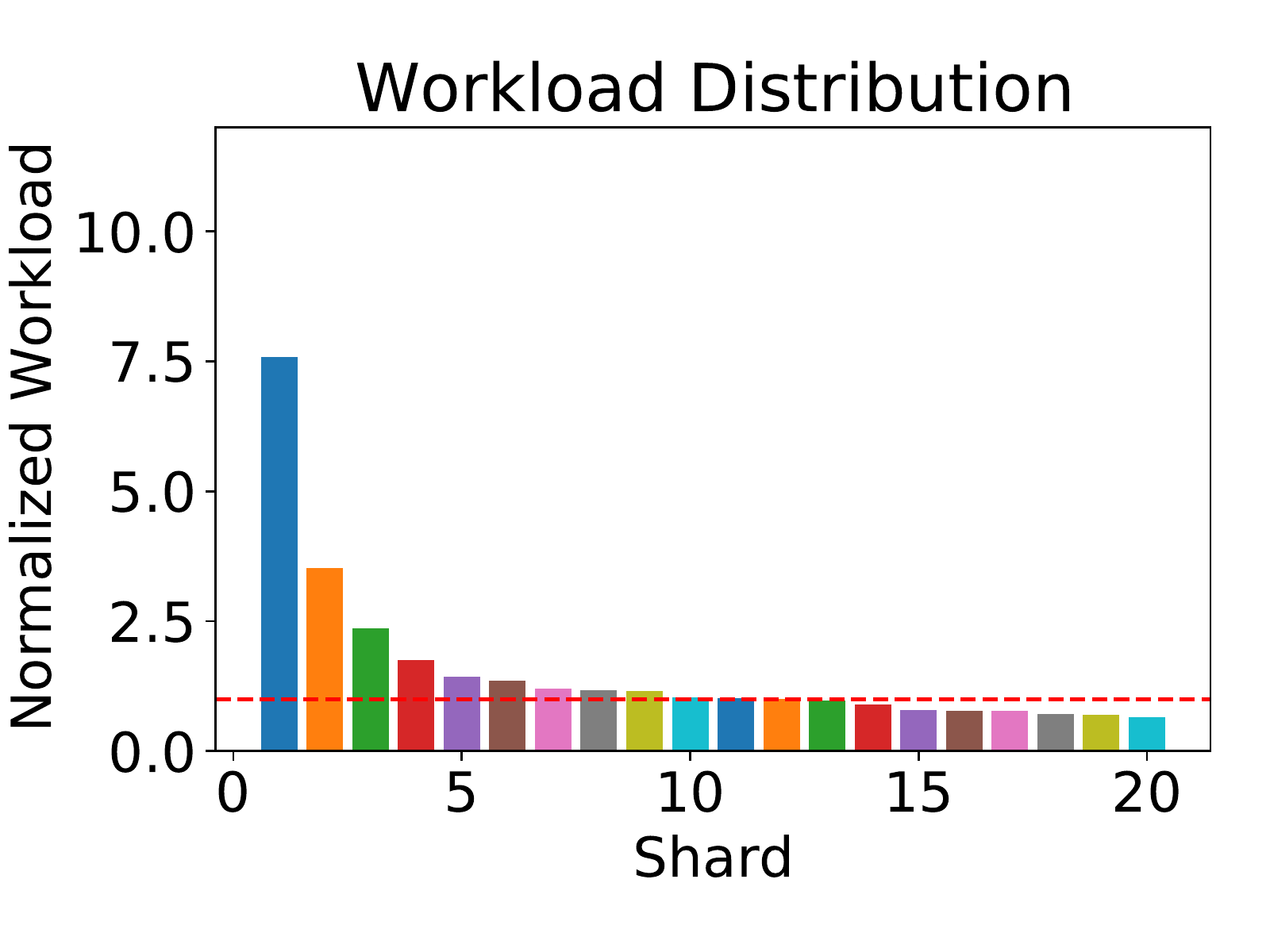}
      \caption{METIS}
       \label{Metisdistribution}
    \end{subfigure}
    \centering
    \begin{subfigure}{0.24\textwidth}
    
     \centering
     \includegraphics[width=1\textwidth]{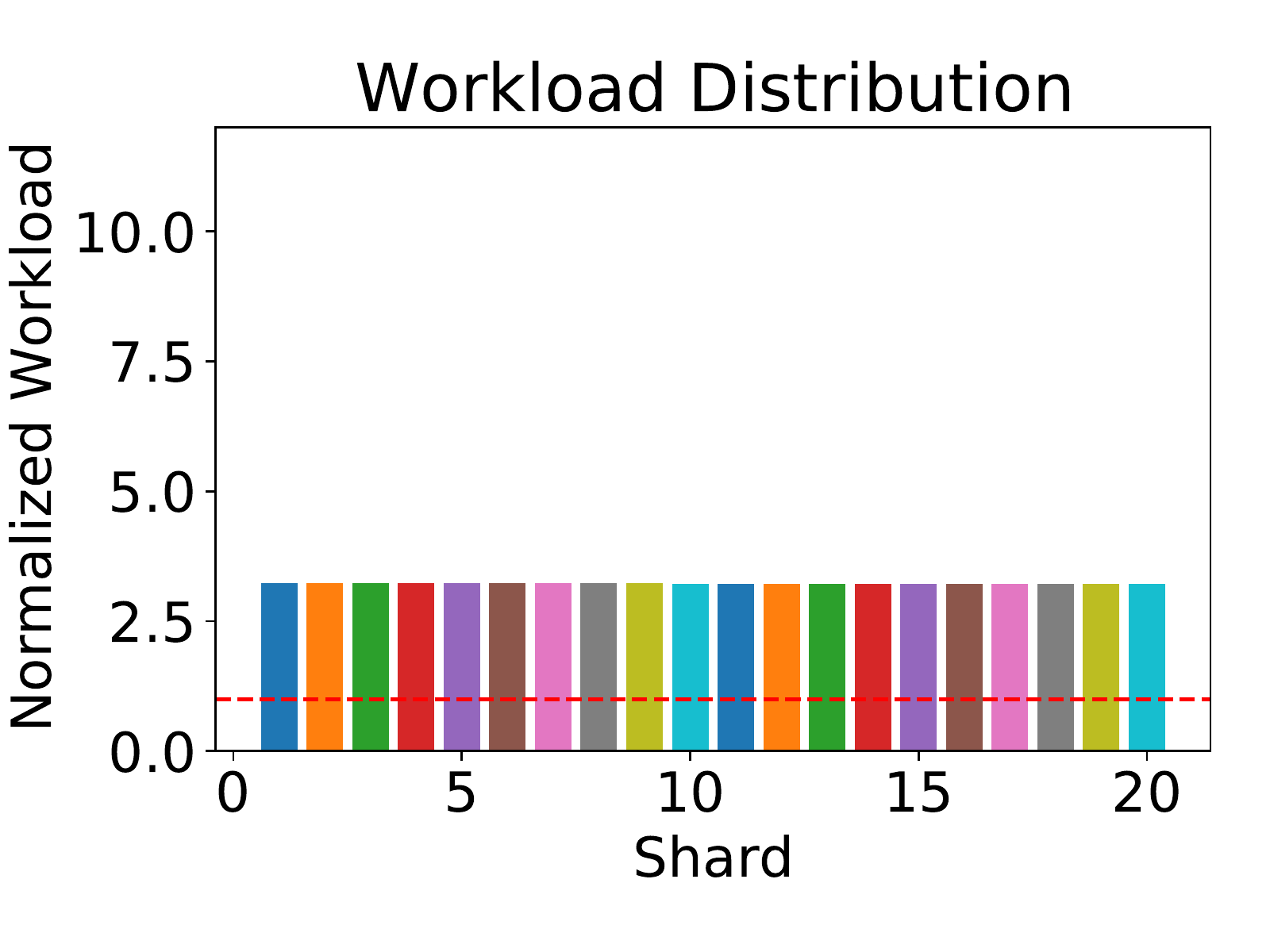}
      \caption{Shard Scheduler}
      \label{Shard Schedulerdistribution}
    \end{subfigure}
    \begin{subfigure}{0.24\textwidth}
     \centering
      \includegraphics[width=1\textwidth]{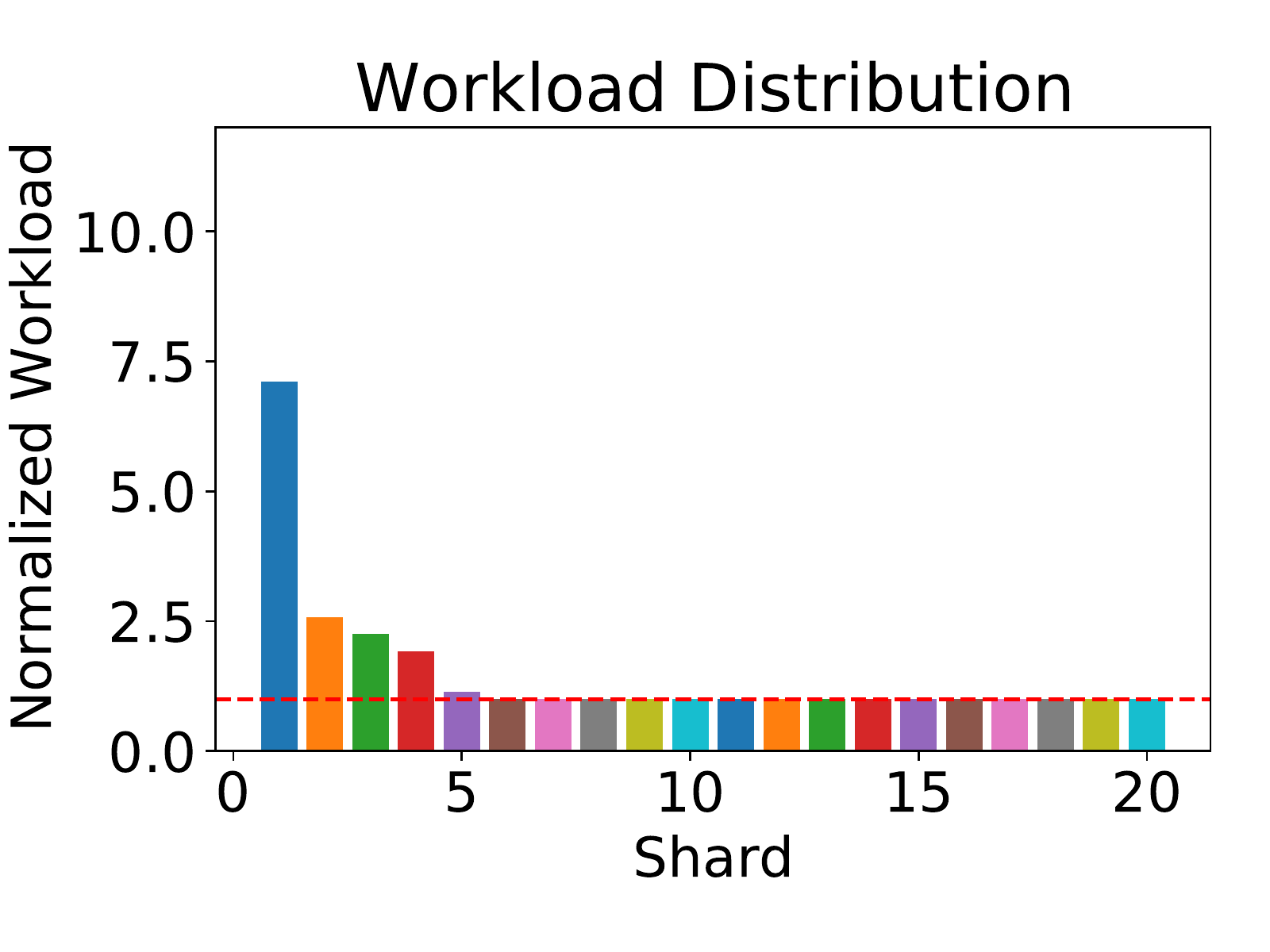}
     \caption{Our method}
      \label{Ourdistribution}
    \end{subfigure}
    \caption{\small A case study example of workload distribution among shards with $\Difficulty =2$ and $k=20$. The y-axis is the normalized workload computed by $\workload_i/\capacity$. The red horizontal line indicates normalized workload is 1, i.e., workload assigned to the shard is exactly equivalent to the processing capacity, $\workload_i=\capacity$.}
    \label{workload distribution}
    \vspace{-5mm}
\end{figure}

\subsection{Ethereum Dataset}
We collect real-world transaction data from Ethereum for experiments~\cite{bigquery}. 
Ethereum is moving towards a sharded blockchain and accumulates massive real-world transactions.
Note that although the experiments are conducted on Ethereum data for a case study, \NAME{} can be applied to other blockchain systems. 
The data includes recent transactions from block 10,000,000 to block 10,600,000 (approximately 3 months from the 5th of May to the 4th of August in 2020).
It contains 91,857,819 transactions and 12,614,390 accounts. 
We expect miners to initialize the G-TxAllo using only recent history rather than the full history, as also recommended in Shard Scheduler. This prevents noise from out-of-date transactions. 
 Figure \ref{vis} illustrates the detailed transaction pattern of this dataset by sampling 300,000 transactions for visualization.
 The large dots representing active accounts and about 11$\%$ transactions are associated with the most active account. Most accounts are not active and only have very few transaction records.
Such a long-tail distributed pattern seriously challenges the balanced allocation in terms of workload.

\subsection{\globalNAME}
\label{sec_global_results}
 We compare \globalNAME{} with the state-of-the-art baselines including the traditional hash-based random allocation~\cite{wang2019monoxide,kokoris2018omniledger,al2017chainspace,zamani2018rapidchain}, METIS-based graph-level allocation~\cite{mizrahi2020blockchain,huang2022brokerchain,fynn2018challenges} and transaction-level allocation method, Shard Scheduler~\cite{krol2021}.  
 As defined in Section~\ref{sec_definition}, the evaluation metrics include cross-shard transaction ratio $\crossratio$, workload balance $\balancemetrics$, throughput $\throughput$, latency and the execution time.

\subsubsection{Experimental Setting}
For comprehensive results, we conduct experiments with various $\Difficulty$ from 2 to 10 and $\numshard$ from 2 to 60.  
The processing capacity $\capacity$ in the simulations should be reasonably set to avoid over-loaded and under-loaded problems in the entire system. 
In particular, we set $\capacity = {\left|\setalltx\right|}/{\numshard}$ for different $\numshard$.
By this setting, the system throughput is exactly $\left|\setalltx\right|$ in the ideal situation that all transactions are intra-shard and each shard has an equivalent workload. 
The convergence threshold $\convergence$ is set as $10^{-5}*\left|\setalltx\right|$.
For the fairness of comparison to the transaction-level allocation method, we set the same capacity and the buffer ratio as 1 for Shard Scheduler.

\subsubsection{Cross-shard Transaction Ratio Comparison}
The account-shard mapping aims to reduce the number of cross-shard transactions for better system throughput.  Figure~\ref{global_ratio} illustrates the comparison in terms of cross-shard transaction ratio $\crossratio$. 
\globalNAME{} shows the effectiveness of reducing the cross-shard transactions in the system and achieves the best $\crossratio$ in this comparison.
It achieves about $12\%$ cross-shard ratio even when $\numshard = 60$. 

\begin{figure*}[!t]
\captionsetup[subfigure]{justification=centering}
    \centering
    \begin{subfigure}{0.17\textwidth}
     \centering
     \includegraphics[width=1\textwidth]{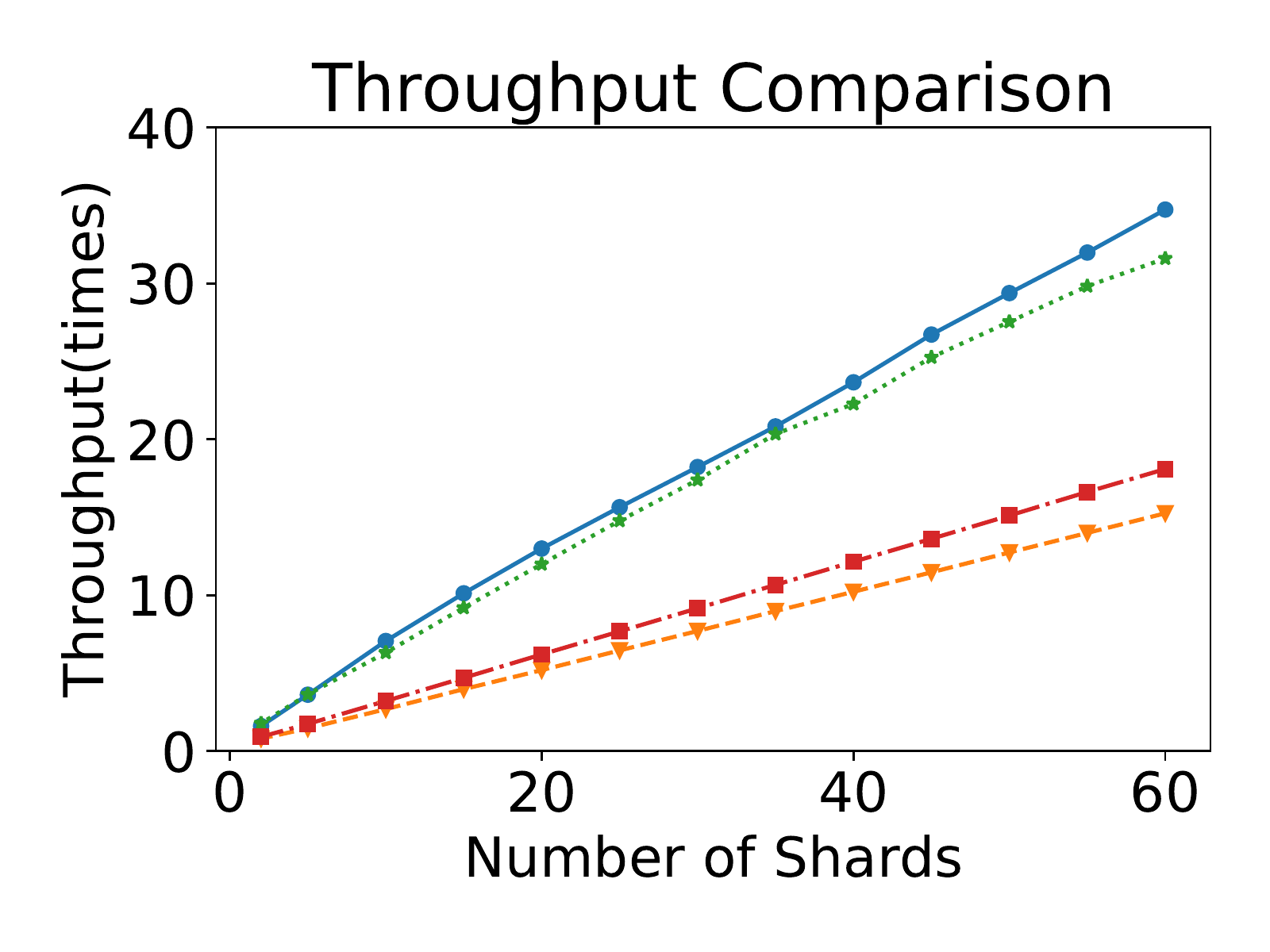}
     \caption{ $\Difficulty = 2$}
    \end{subfigure}
    \begin{subfigure}{0.17\textwidth}
     \centering
      \includegraphics[width=1\textwidth]{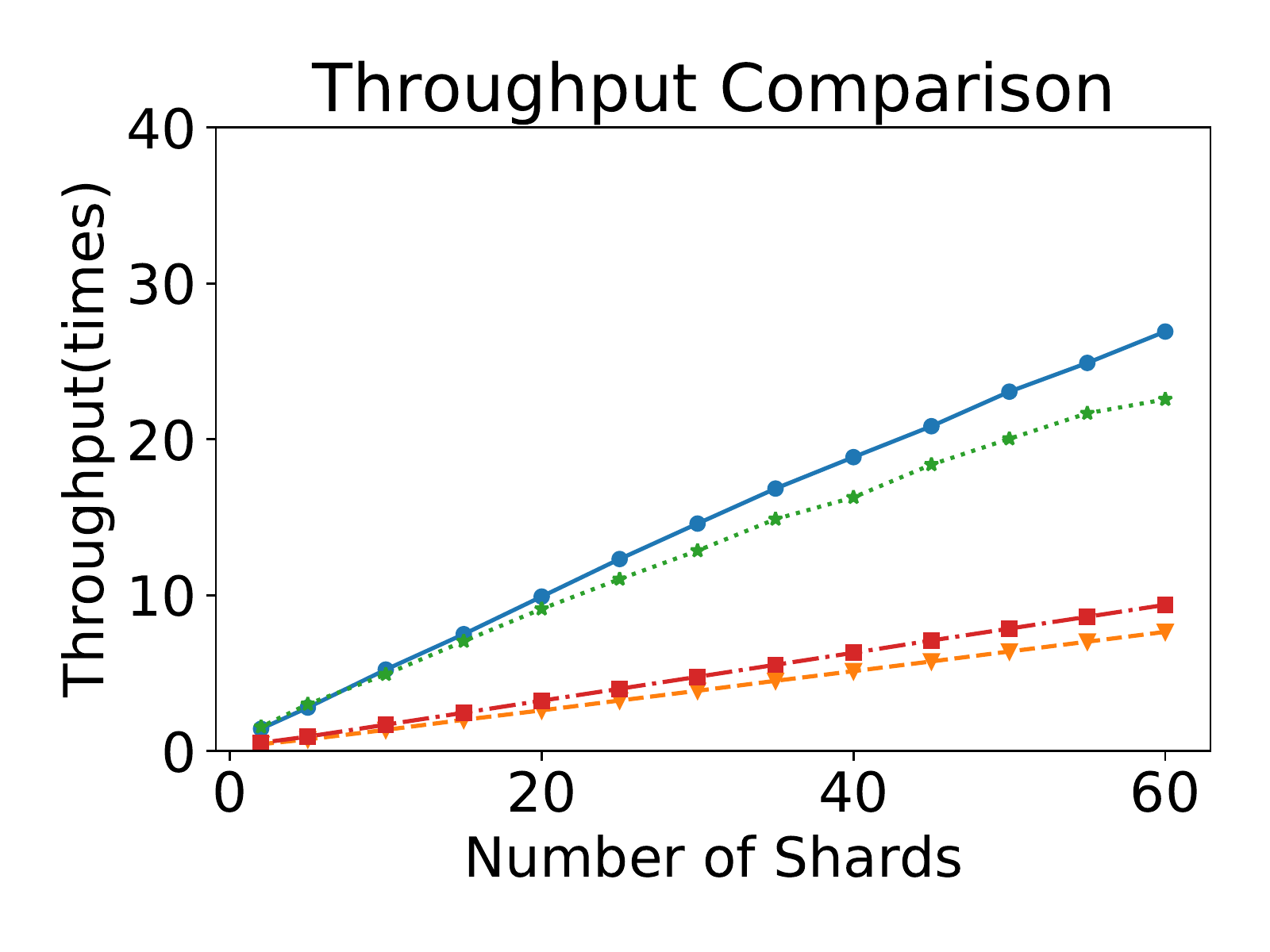}
      \caption{ $\Difficulty = 4$}
    \end{subfigure}
    \centering
    \begin{subfigure}{0.17\textwidth}
     \centering
     \includegraphics[width=1\textwidth]{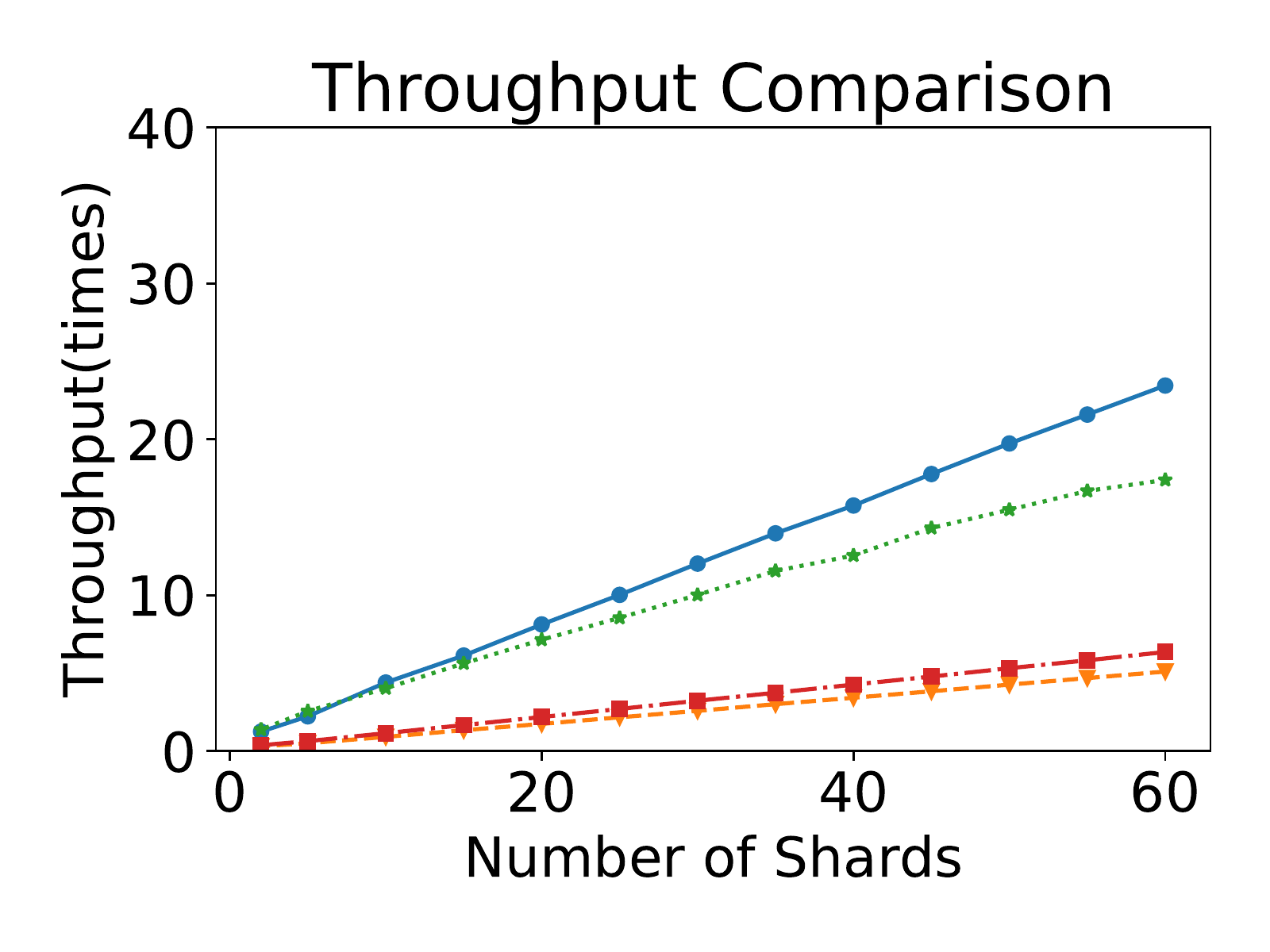}
     \caption{ $\Difficulty = 6$}
    \end{subfigure}
    \begin{subfigure}{0.17\textwidth}
     \centering
      \includegraphics[width=1\textwidth]{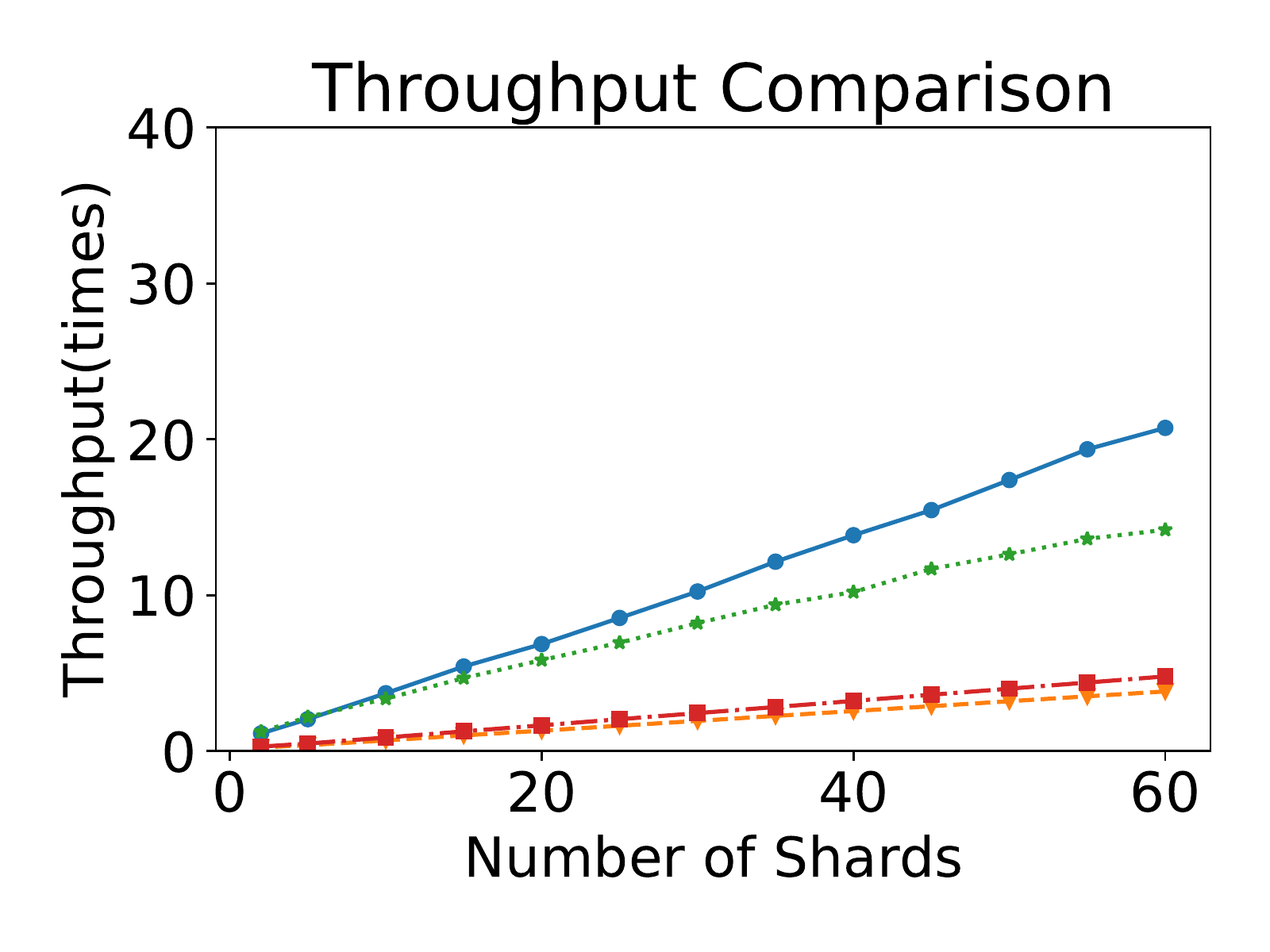}
      \caption{ $\Difficulty = 8$}
    \end{subfigure}
    \begin{subfigure}{0.26\textwidth}
     \centering
      \includegraphics[width=1\textwidth]{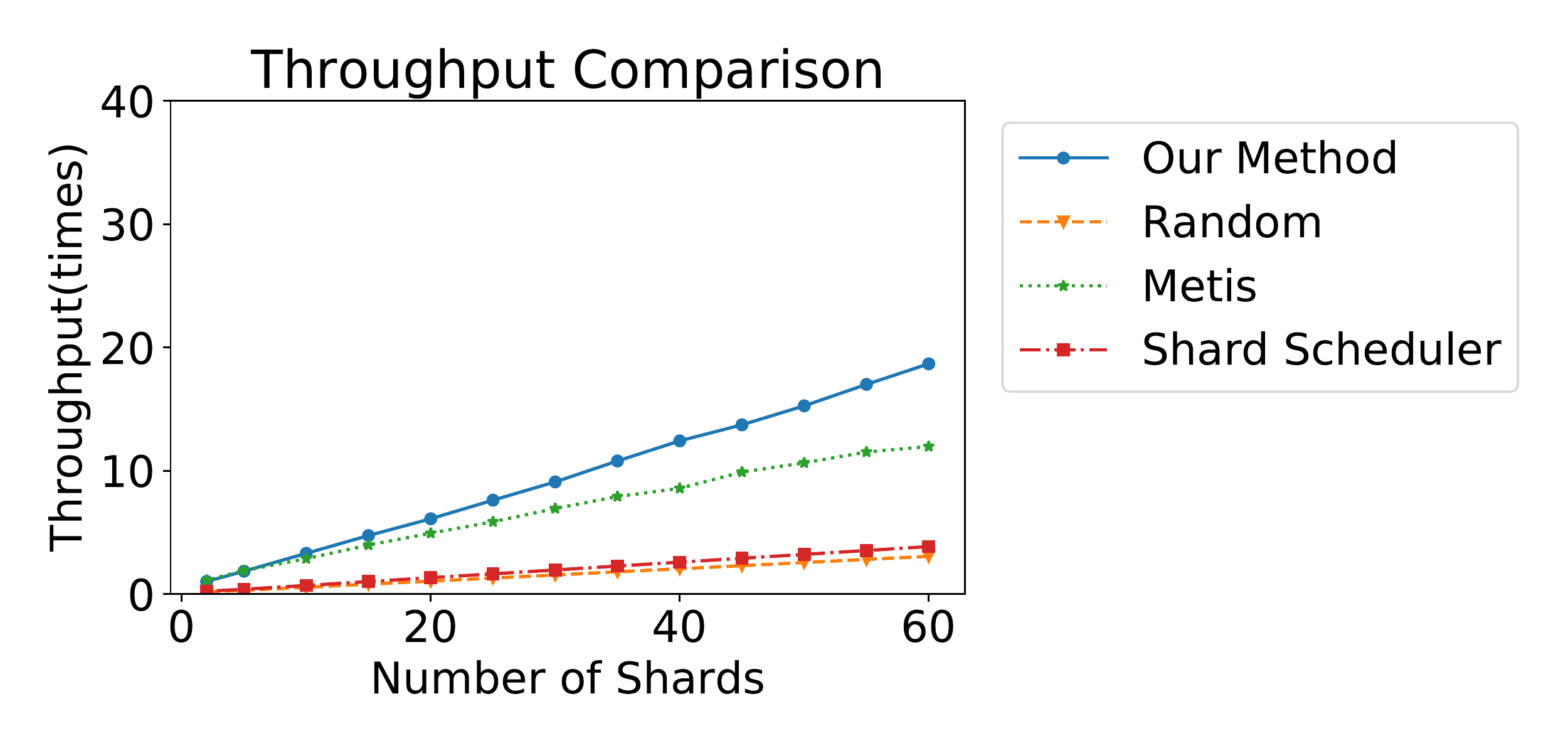}
      \caption{ $\Difficulty = 10$}
    \end{subfigure}
    \caption{Throughput comparison with various $\numrelashards$ and $\Difficulty$.  The x-axis is the number of shards and the y-axis indicates how many times the throughput is improved from an unsharded blockchain.}
    \label{global_throughput}
\end{figure*}

\begin{figure*}[ht]
\captionsetup[subfigure]{justification=centering}
    \centering
    \begin{subfigure}{0.17\textwidth}
     \centering
     \includegraphics[width=1\textwidth]{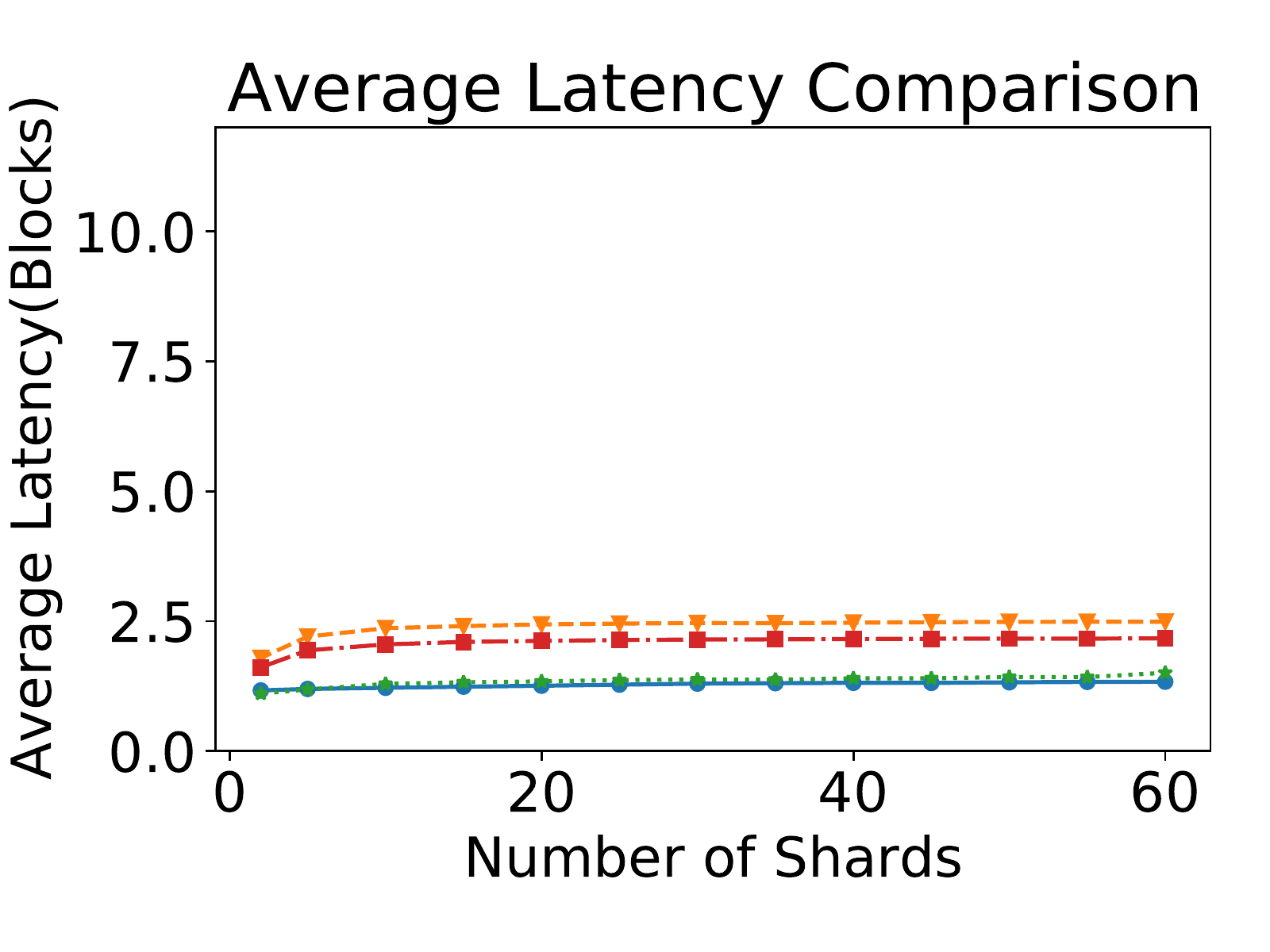}
     \caption{ $\Difficulty = 2$}
    \end{subfigure}
    \begin{subfigure}{0.17\textwidth}
     \centering
      \includegraphics[width=1\textwidth]{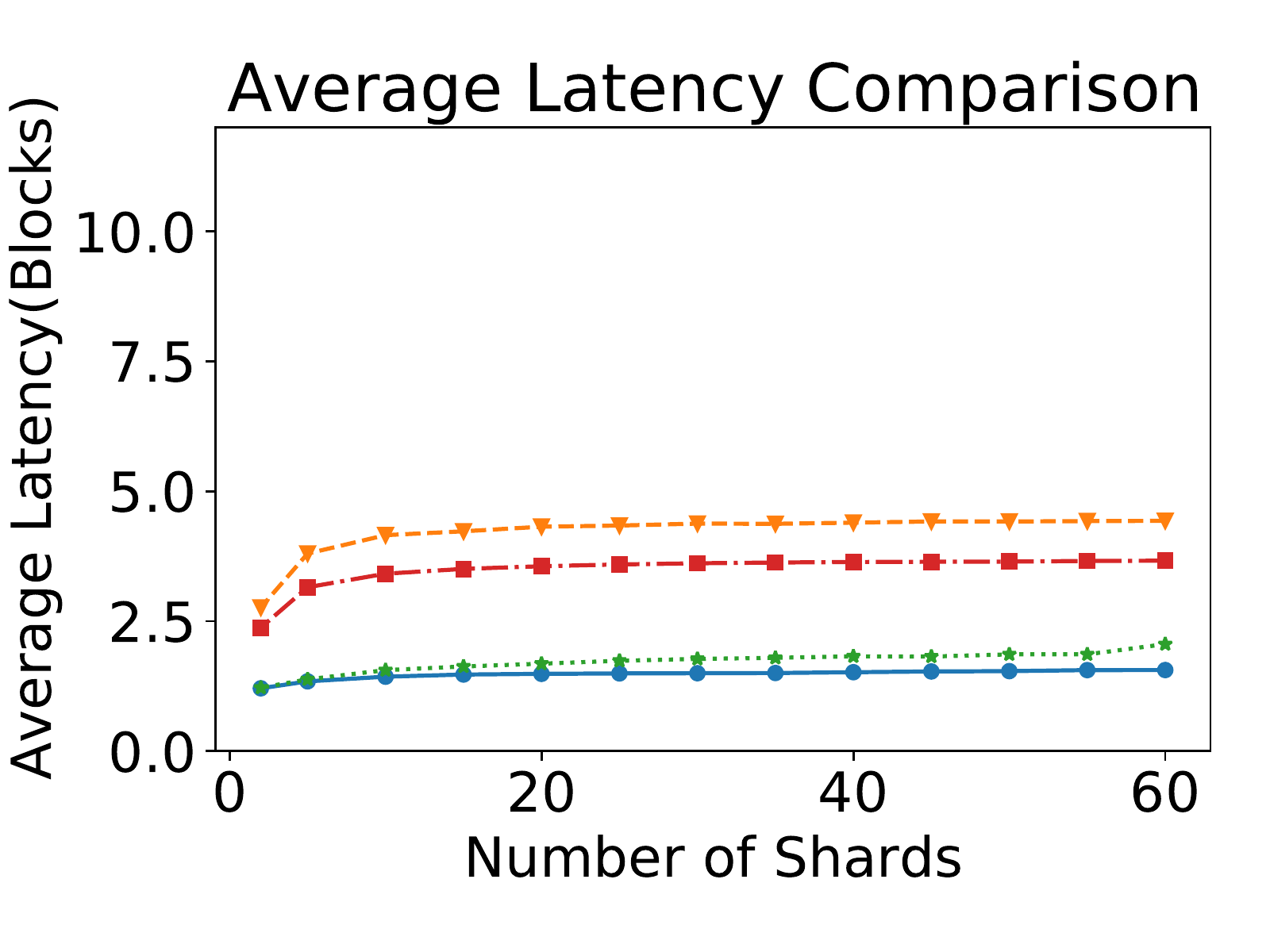}
      \caption{ $\Difficulty = 4$}
    \end{subfigure}
    \centering
    \begin{subfigure}{0.17\textwidth}
     \centering
     \includegraphics[width=1\textwidth]{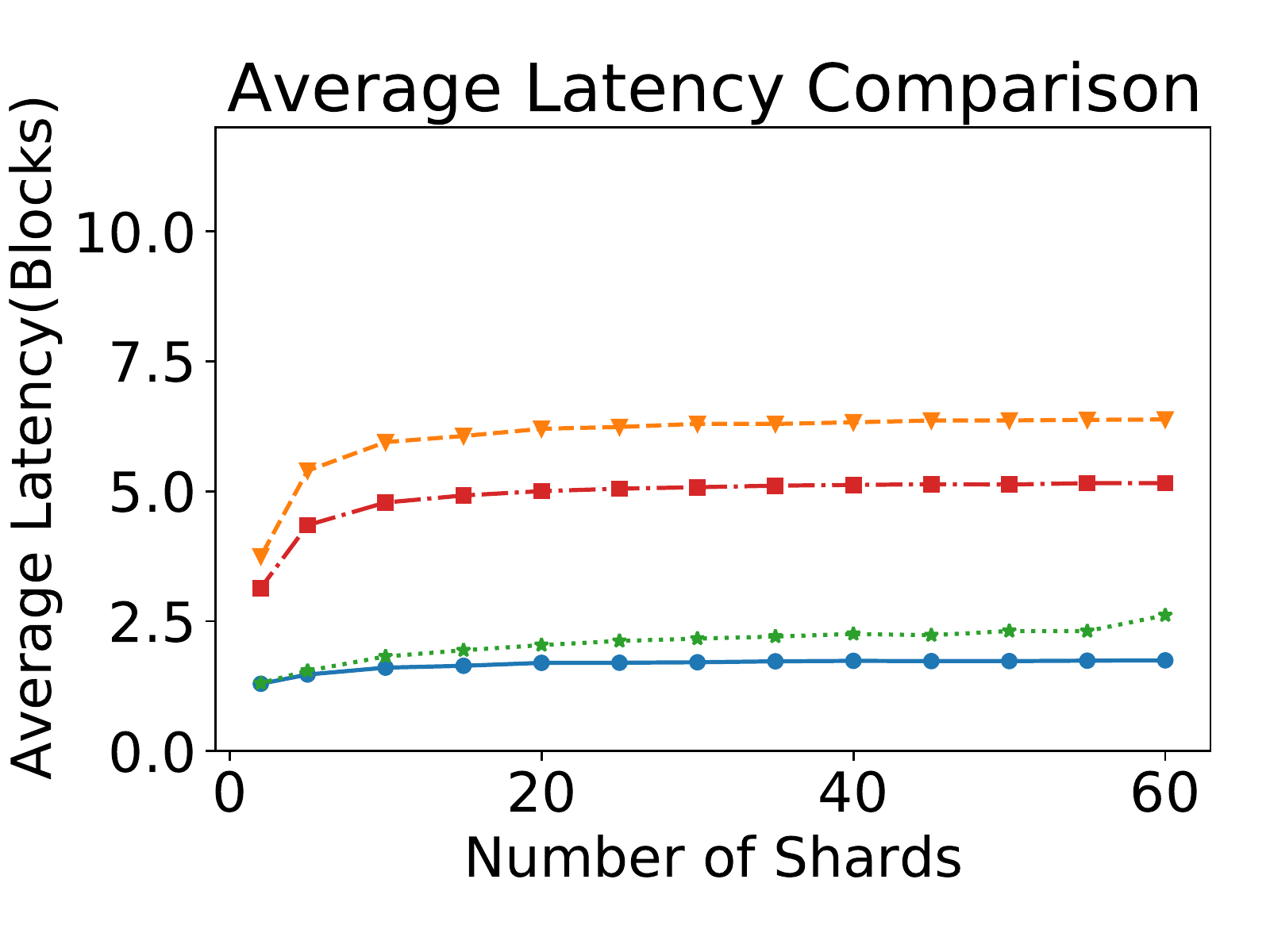}
     \caption{ $\Difficulty = 6$}
    \end{subfigure}
    \begin{subfigure}{0.17\textwidth}
     \centering
      \includegraphics[width=1\textwidth]{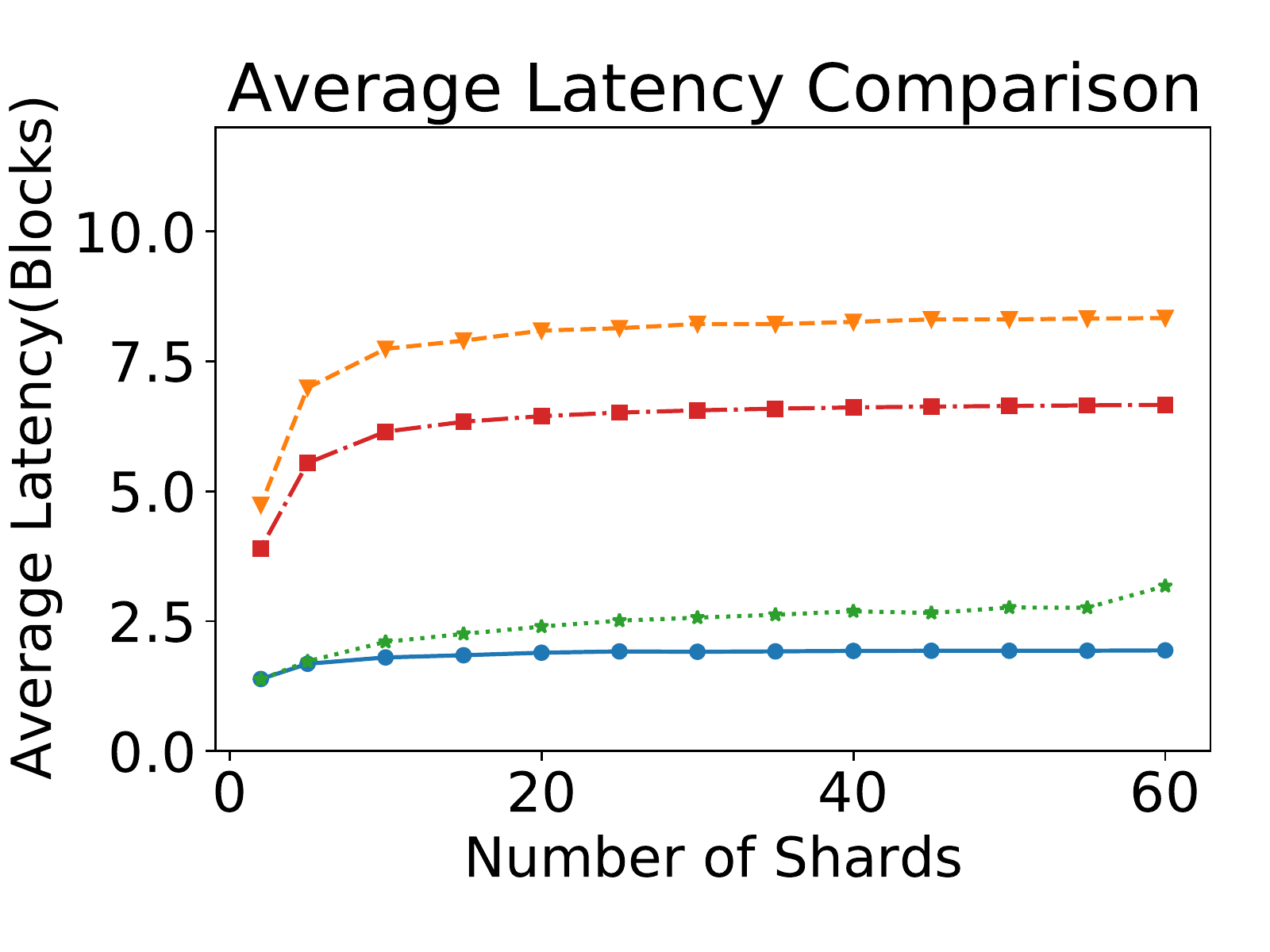}
      \caption{ $\Difficulty = 8$}
    \end{subfigure}
    \begin{subfigure}{0.26\textwidth}
     \centering
      \includegraphics[width=1\textwidth]{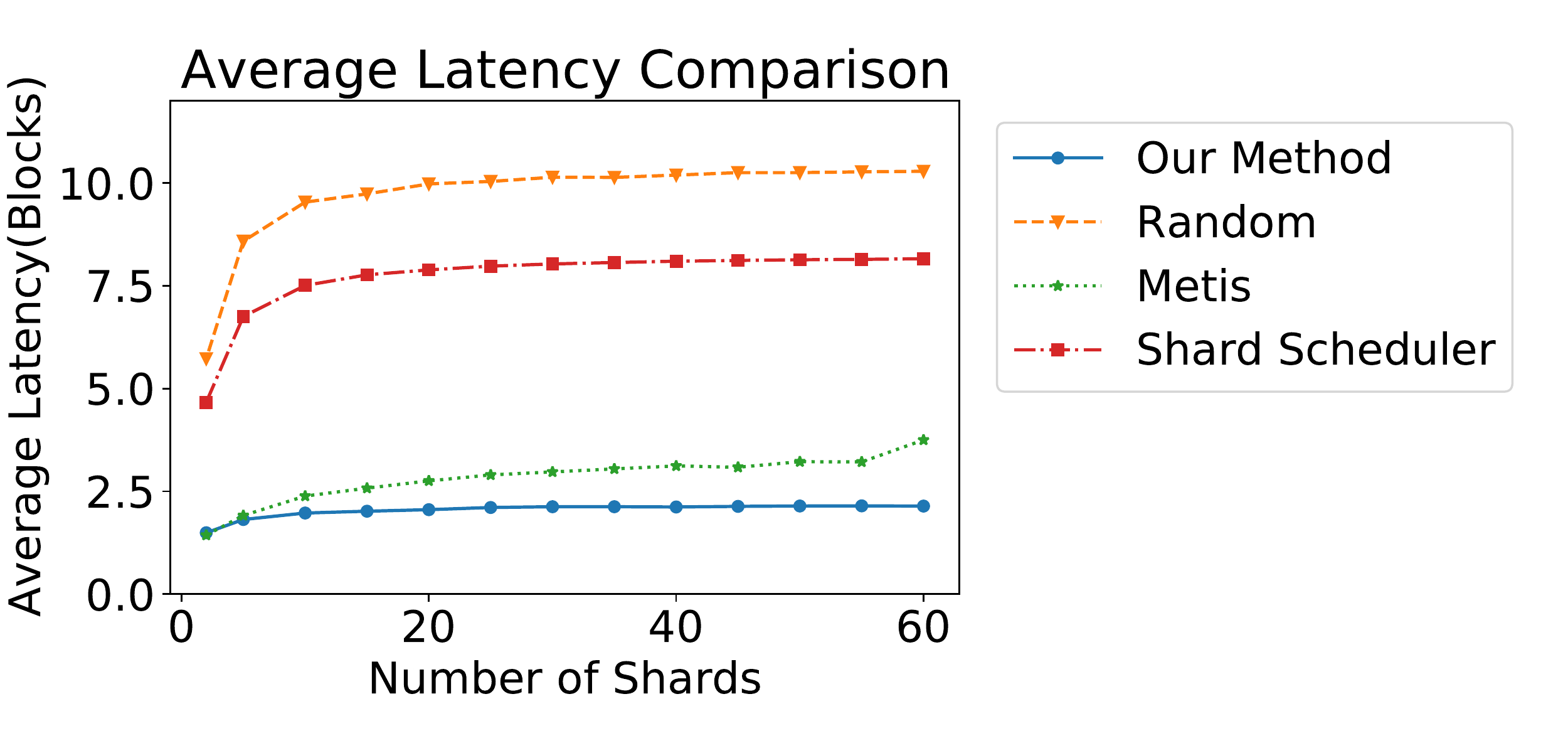}
      \caption{ $\Difficulty = 10$}
    \end{subfigure}
    \caption{Average latency comparison with various $\numrelashards$ and $\Difficulty$. }
    \label{global_average_latency}
\end{figure*}

\begin{figure*}[ht]
    \captionsetup[subfigure]{justification=centering}
    \centering
    \begin{subfigure}{0.17\textwidth}
     \centering
     \includegraphics[width=1\textwidth]{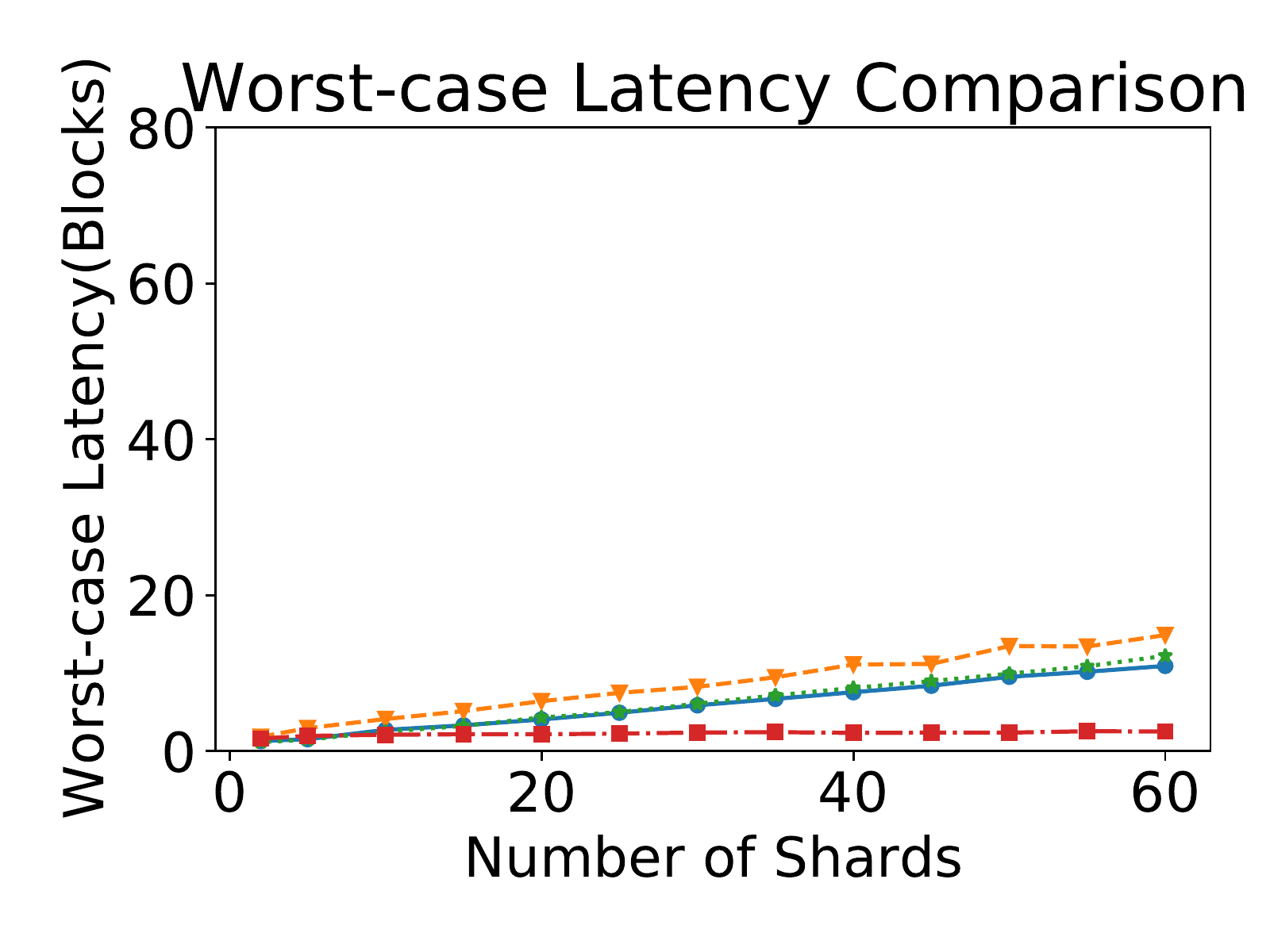}
     \caption{ $\Difficulty = 2$}
    \end{subfigure}
    \begin{subfigure}{0.17\textwidth}
     \centering
      \includegraphics[width=1\textwidth]{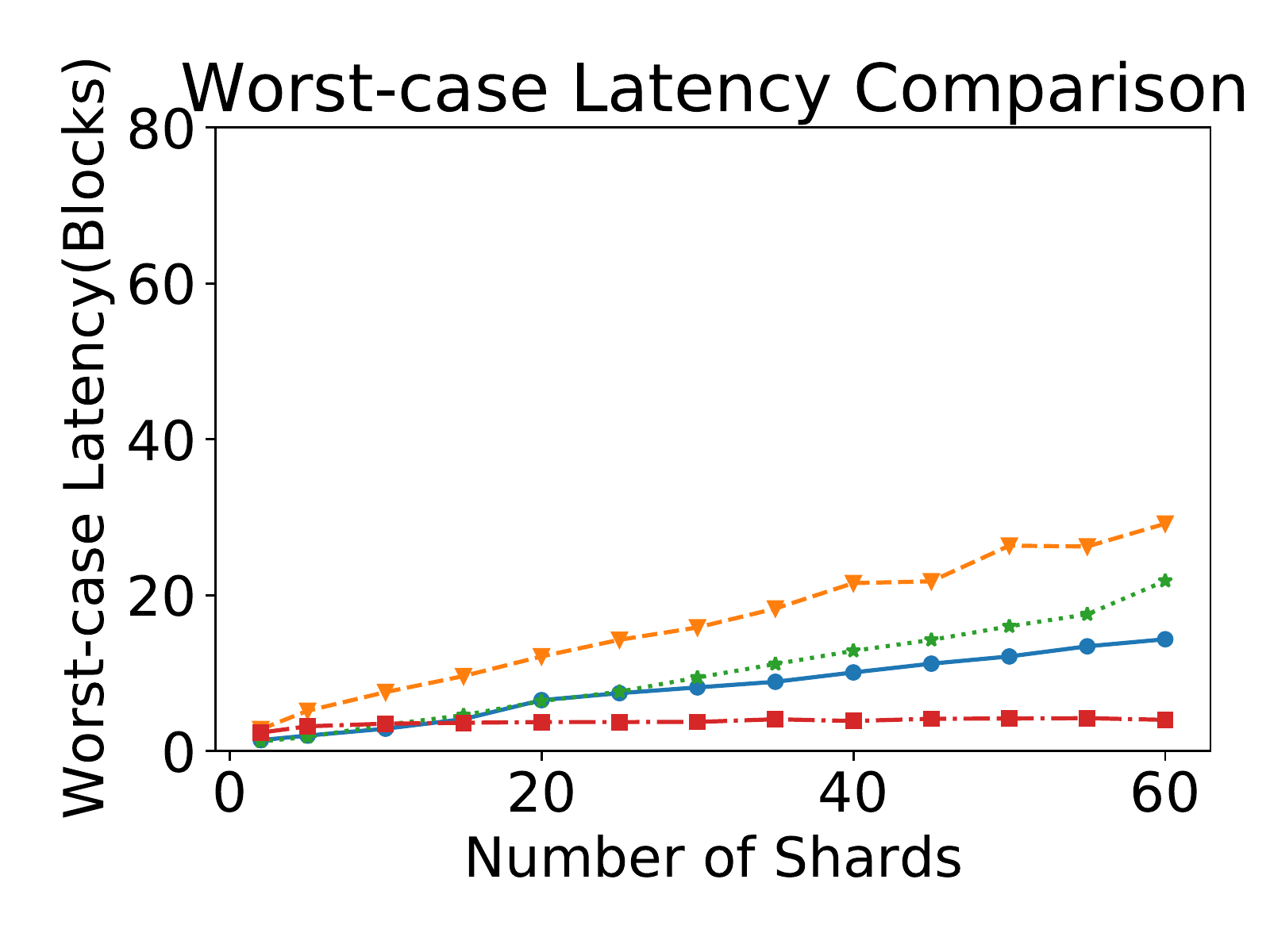}
      \caption{ $\Difficulty = 4$}
    \end{subfigure}
    \centering
    \begin{subfigure}{0.17\textwidth}
     \centering
     \includegraphics[width=1\textwidth]{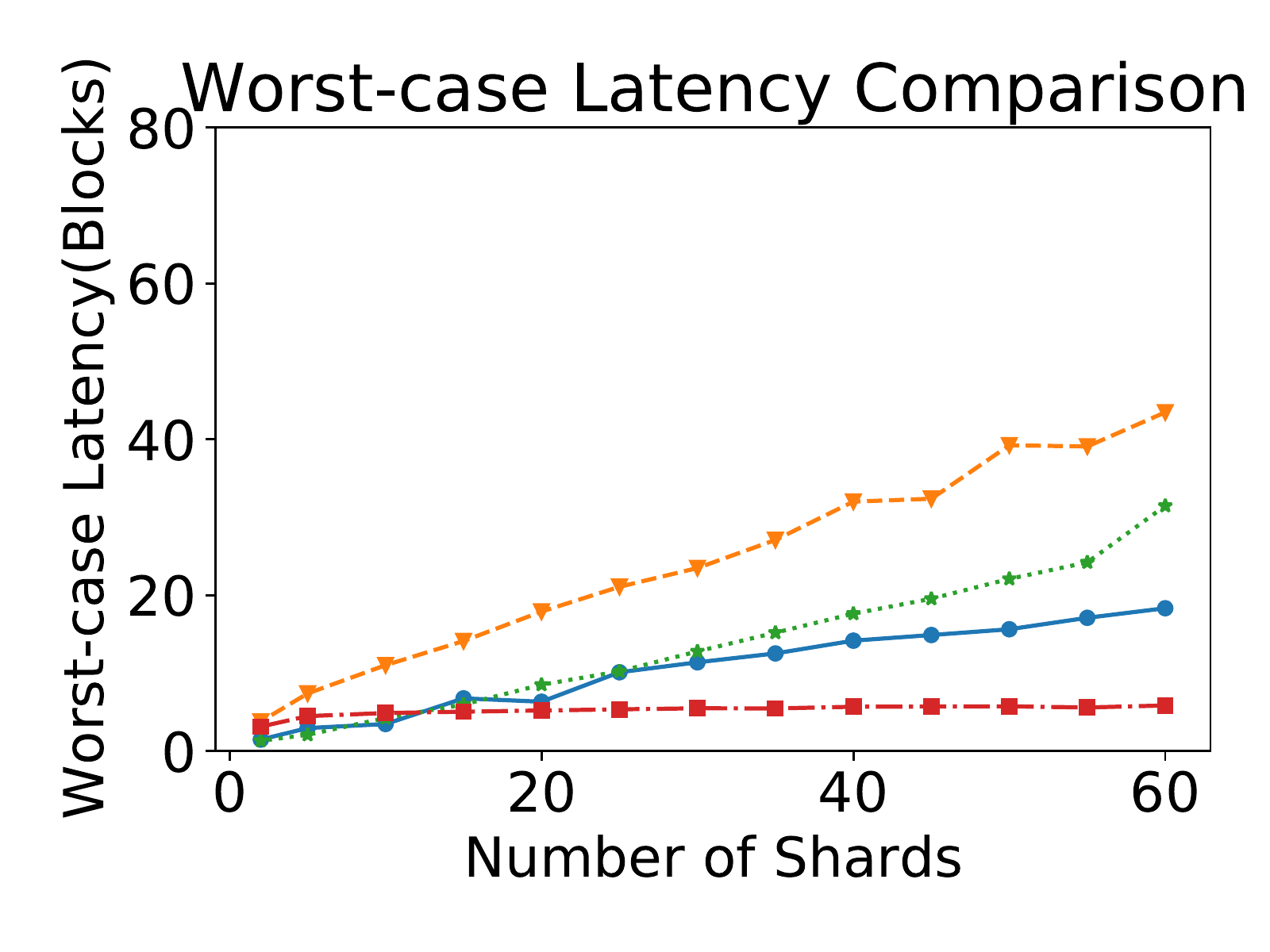}
     \caption{ $\Difficulty = 6$}
    \end{subfigure}
    \begin{subfigure}{0.17\textwidth}
     \centering
      \includegraphics[width=1\textwidth]{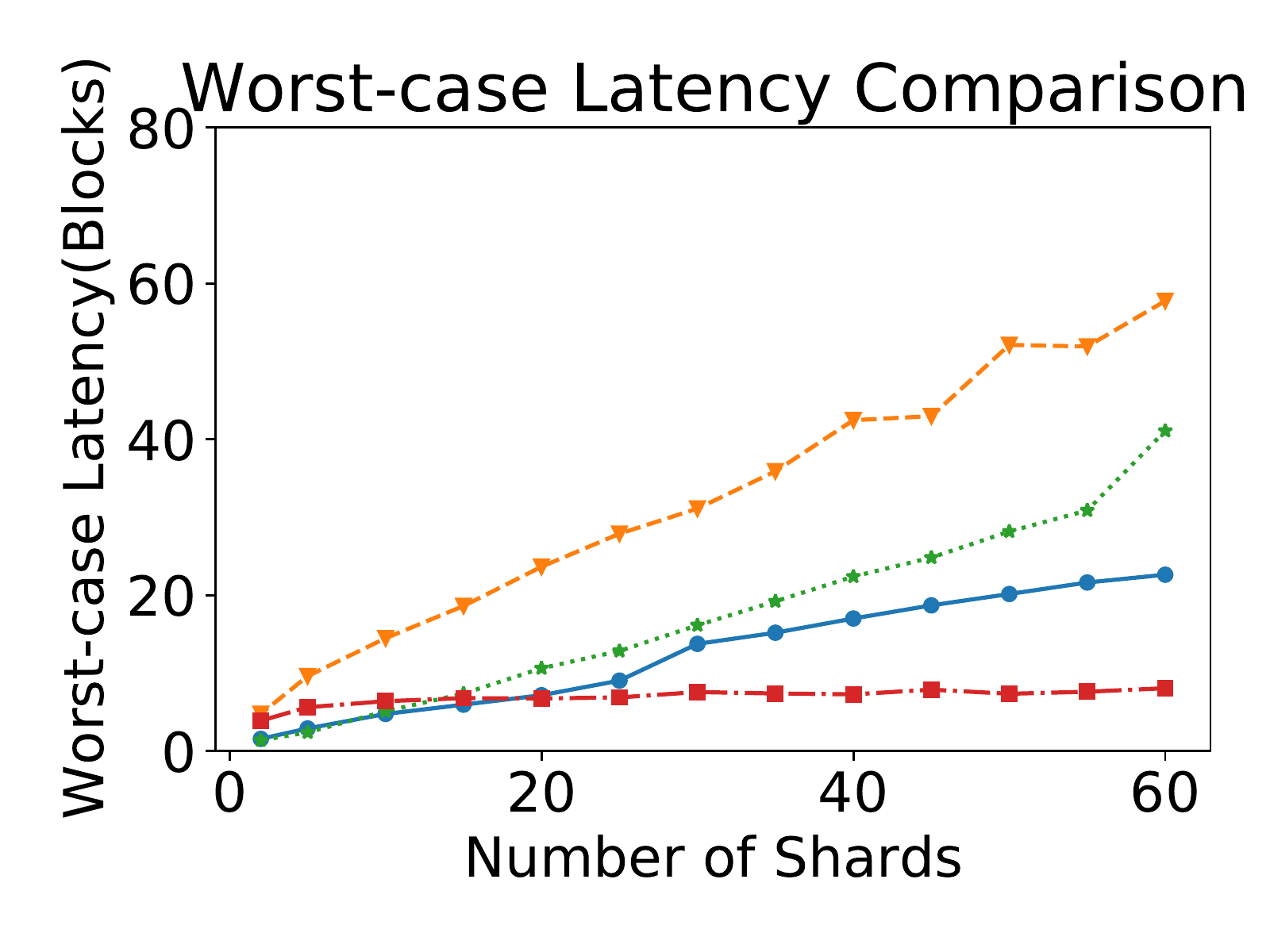}
      \caption{ $\Difficulty = 8$}
    \end{subfigure}
    \begin{subfigure}{0.26\textwidth}
     \centering
      \includegraphics[width=1\textwidth]{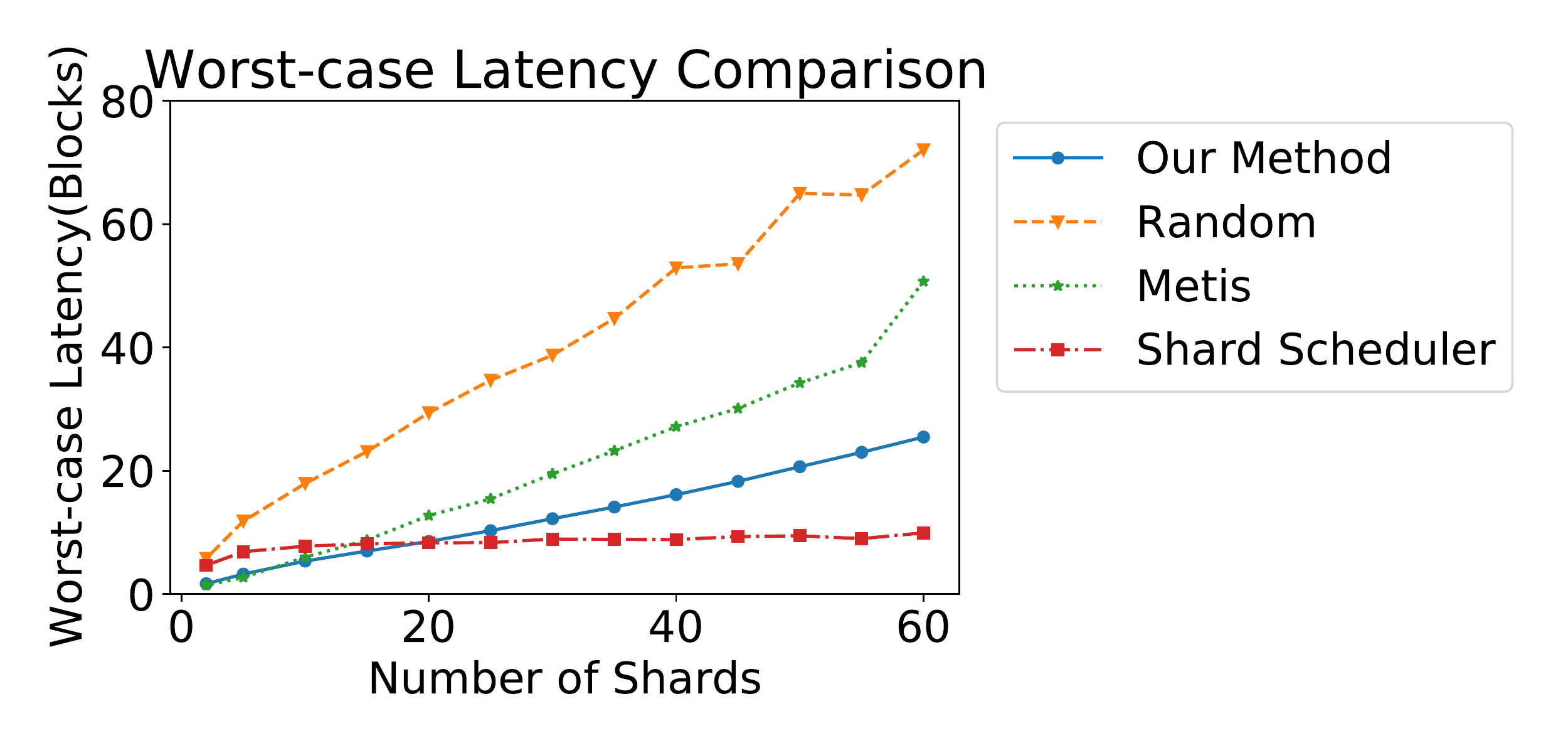}
      \caption{ $\Difficulty = 10$}
    \end{subfigure}
    \caption{Worst-case latency comparison with various $\numrelashards$ and $\Difficulty$. }
    \label{global_worst-case_latency}
\end{figure*}

 \globalNAME{} can self-adjust the cross-shard ratio for different $\Difficulty$ as it directly optimizes throughput $\throughput$ to implicitly optimizes $\crossratio$.
 When $\Difficulty$ is large, the system throughput $\throughput$ is influenced more by cross-shard transactions and \globalNAME{} prioritizes to optimize $\crossratio$ for better $\throughput$.
 This nature of self-adjusting is novel, making \globalNAME{} applicable to different application scenarios.
 The METIS method optimizes its own objectives, e.g. balance of node weight in each community, without considering the blockchain concept: workload.

\subsubsection{Workload Balance Comparison}
The workload balance for account-shard mapping is challenging because of the imbalanced transaction pattern, as shown in Figure~\ref{vis}.

Figure~\ref{global_balance} shows the comparison of workload balance metric $\balancemetrics$. 
 The transaction-level method, Shard Scheduler, achieves the best results and \globalNAME{} is better than the other two graph-based baselines. 
 Figure~\ref{workload distribution} shows the detailed workload distribution among shards for comparison.
Figure~\ref{Randomdistribution} has the most of the total workload as the random method introduces the most cross-shard transactions.
In Figure~\ref{Randomdistribution},~\ref{Metisdistribution} and~\ref{Ourdistribution}, the most over-loaded shard stands out as the most active account with 11\% total transactions is allocated in this shard.
Shard Scheduler in Figure~\ref{Shard Schedulerdistribution} avoids this problem through transaction-level allocation and achieves the best balanced workload among shards.
The last several shards in Figure~\ref{Metisdistribution} of the METIS method have the workload below the horizontal line, indicating that these shards do not fully leverage their processing capacity.

\subsubsection{Throughput Comparison}
 We normalize the throughput as ${\throughput}/{\capacity}$ for different $k$. In this way, the throughput of a non-sharded blockchain (i.e. when $k=1$) is normalized to 1, enabling a clear illustration on how many times the throughput is improved with different $k$.  
 
Figure~\ref{global_throughput} illustrates the comparison in terms of throughput with baselines.
Throughput is improved linearly with the number of shards $k$ for all methods and \globalNAME{} achieves the fastest increasing speed.
When $\Difficulty$ is getting larger, due to the existence of cross-shard transactions, the throughput of all methods decreases.
\globalNAME{} is the most stable as it achieves the lowest $\crossratio$ and can be self-adjustable with large $\Difficulty$.

\globalNAME{} and the METIS method achieve similar $\throughput$ when $\Difficulty$ is small.
For example,  when $\Difficulty=2$ and $\numrelashards =60$, \globalNAME{} and METIS achieve $34.7$ times and $31.6$ times throughput improvement, respectively, which indicates about $10\%$ difference.
This is reasonable as the equation~\ref{equ_workloadinshard} of workload $\workload_i$  degenerates to the sum of degrees in this shard when $\Difficulty=1$.
In this situation, the classic method METIS happens to optimize the workload balance.

\begin{figure*}[ht]
\captionsetup[subfigure]{justification=centering}
    \centering
    \begin{subfigure}{0.17\textwidth}
     \centering
     \includegraphics[width=1\textwidth]{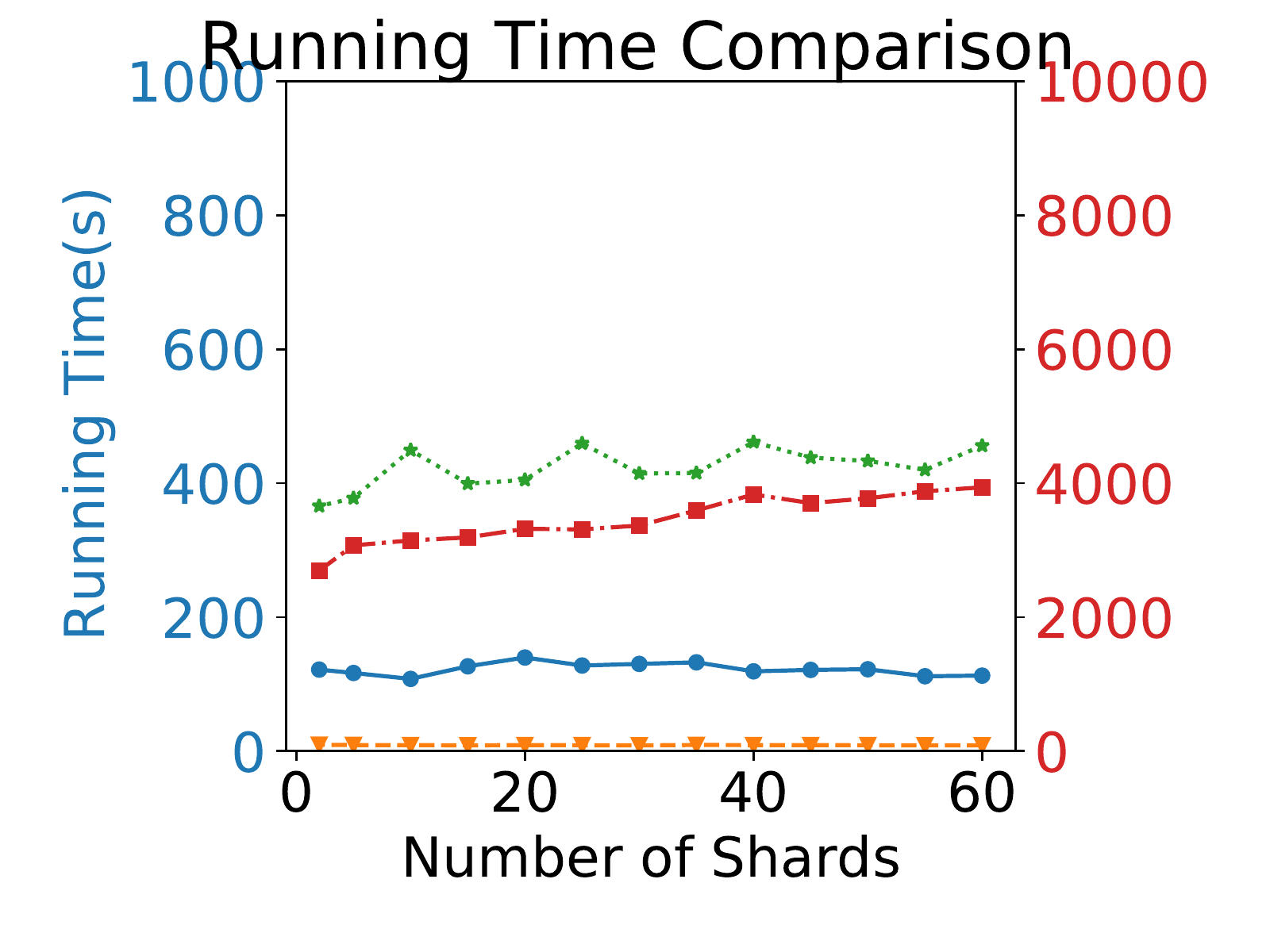}
     \caption{ $\Difficulty = 2$}
    \end{subfigure}
    \begin{subfigure}{0.17\textwidth}
     \centering
      \includegraphics[width=1\textwidth]{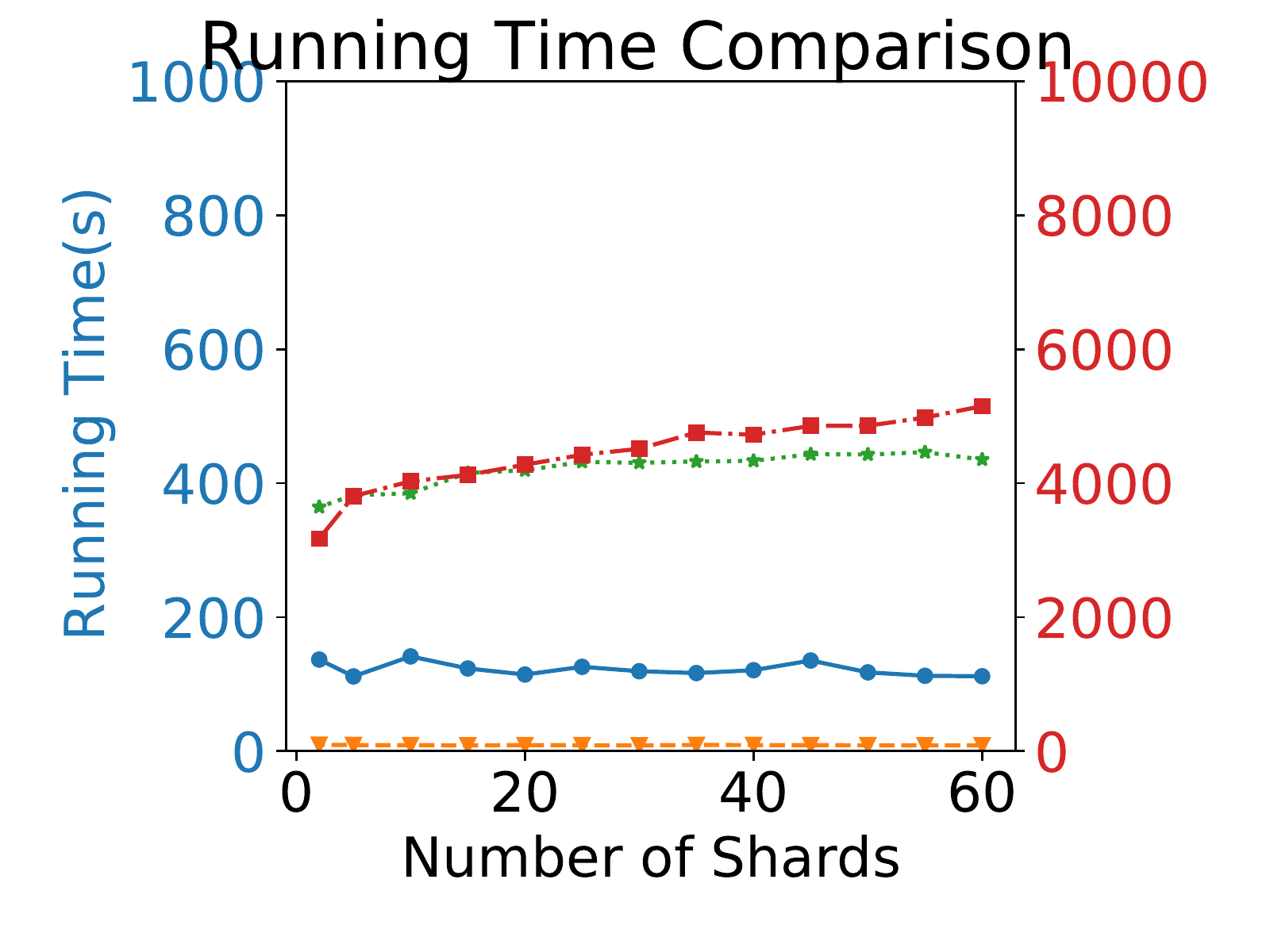}
      \caption{ $\Difficulty = 4$}
    \end{subfigure}
    \centering
    \begin{subfigure}{0.17\textwidth}
     \centering
     \includegraphics[width=1\textwidth]{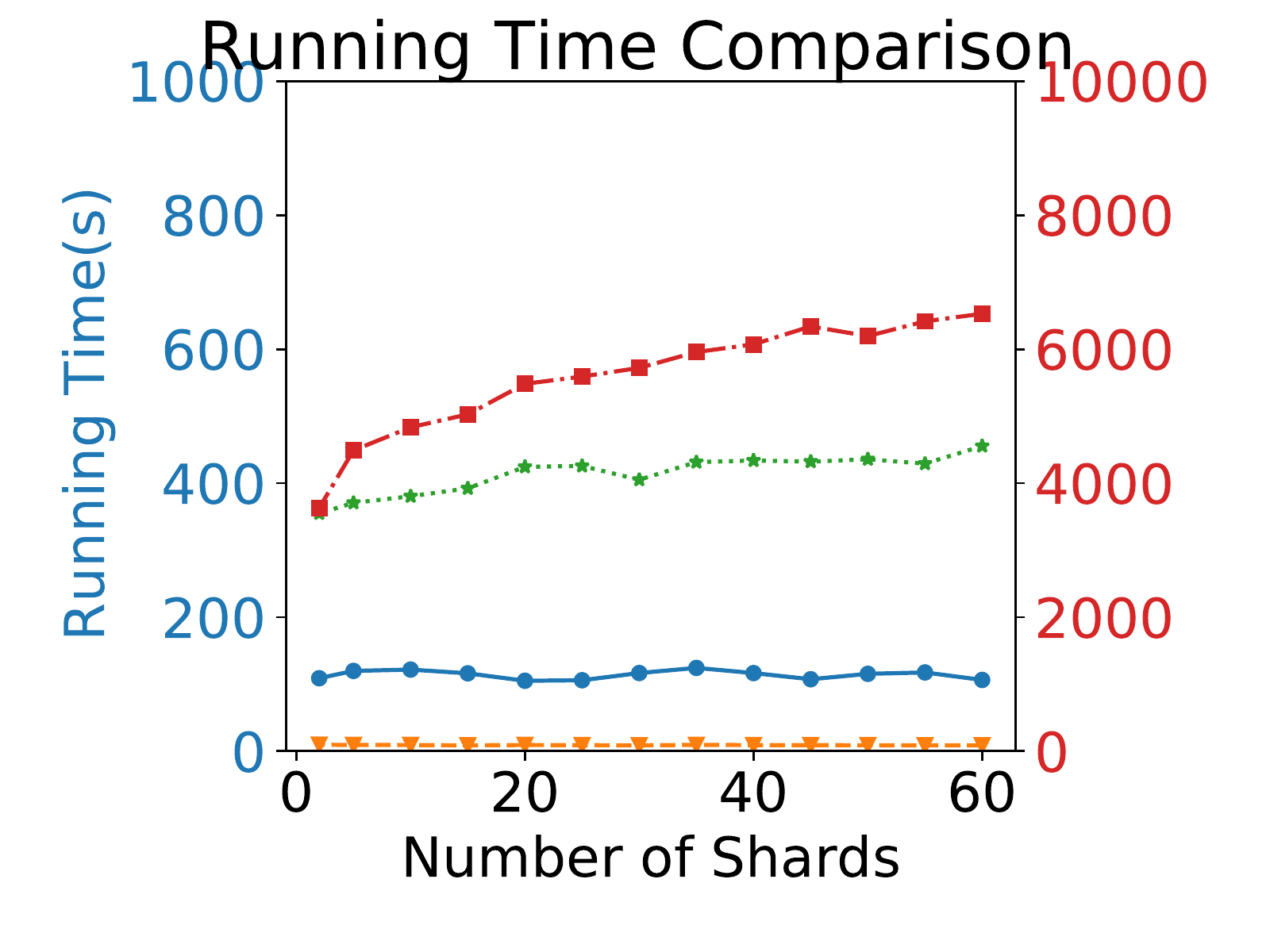}
     \caption{ $\Difficulty = 6$}
    \end{subfigure}
    \begin{subfigure}{0.17\textwidth}
     \centering
      \includegraphics[width=1\textwidth]{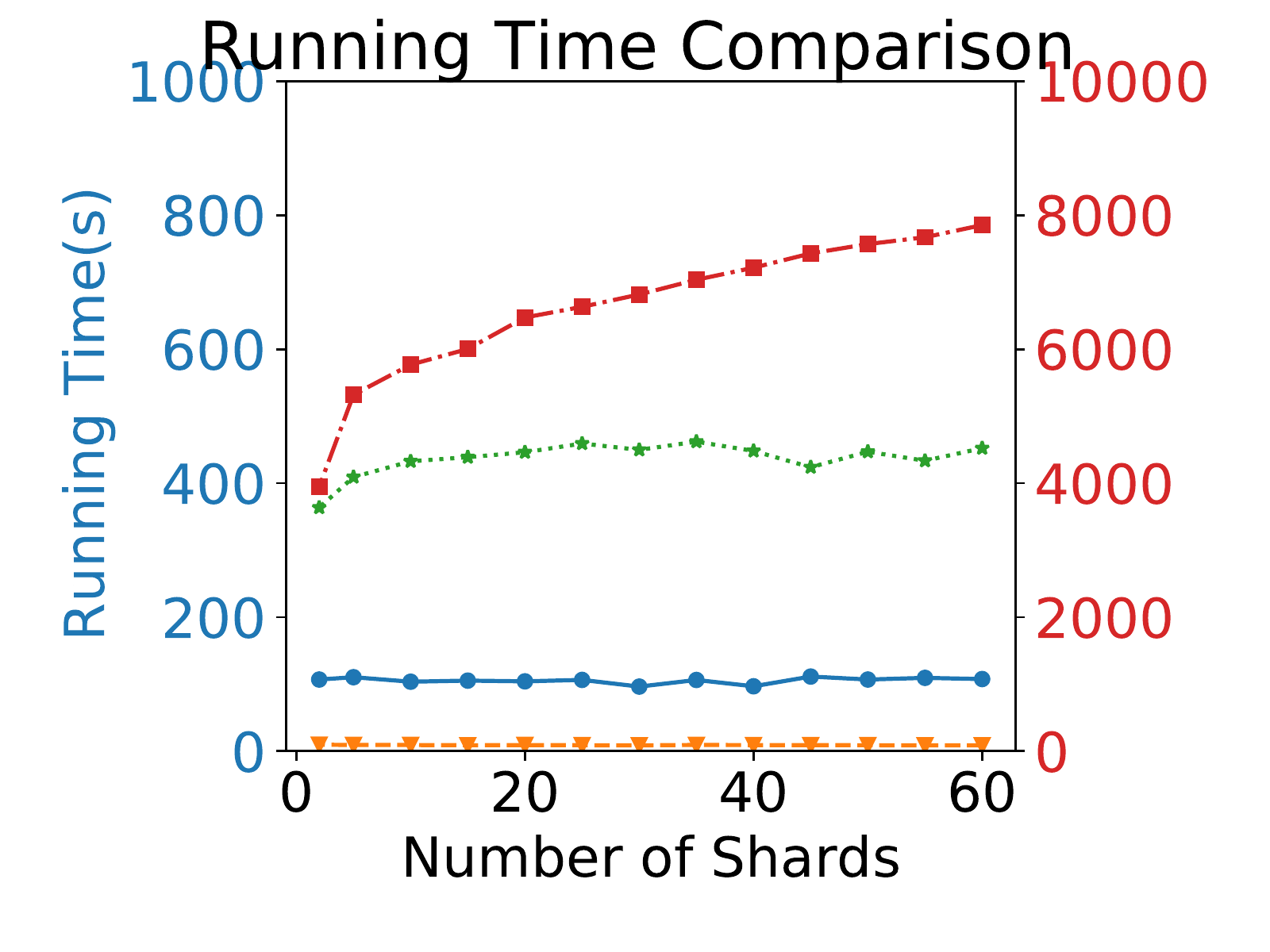}
      \caption{ $\Difficulty = 8$}
    \end{subfigure}
    \begin{subfigure}{0.26\textwidth}
     \centering
      \includegraphics[width=1\textwidth]{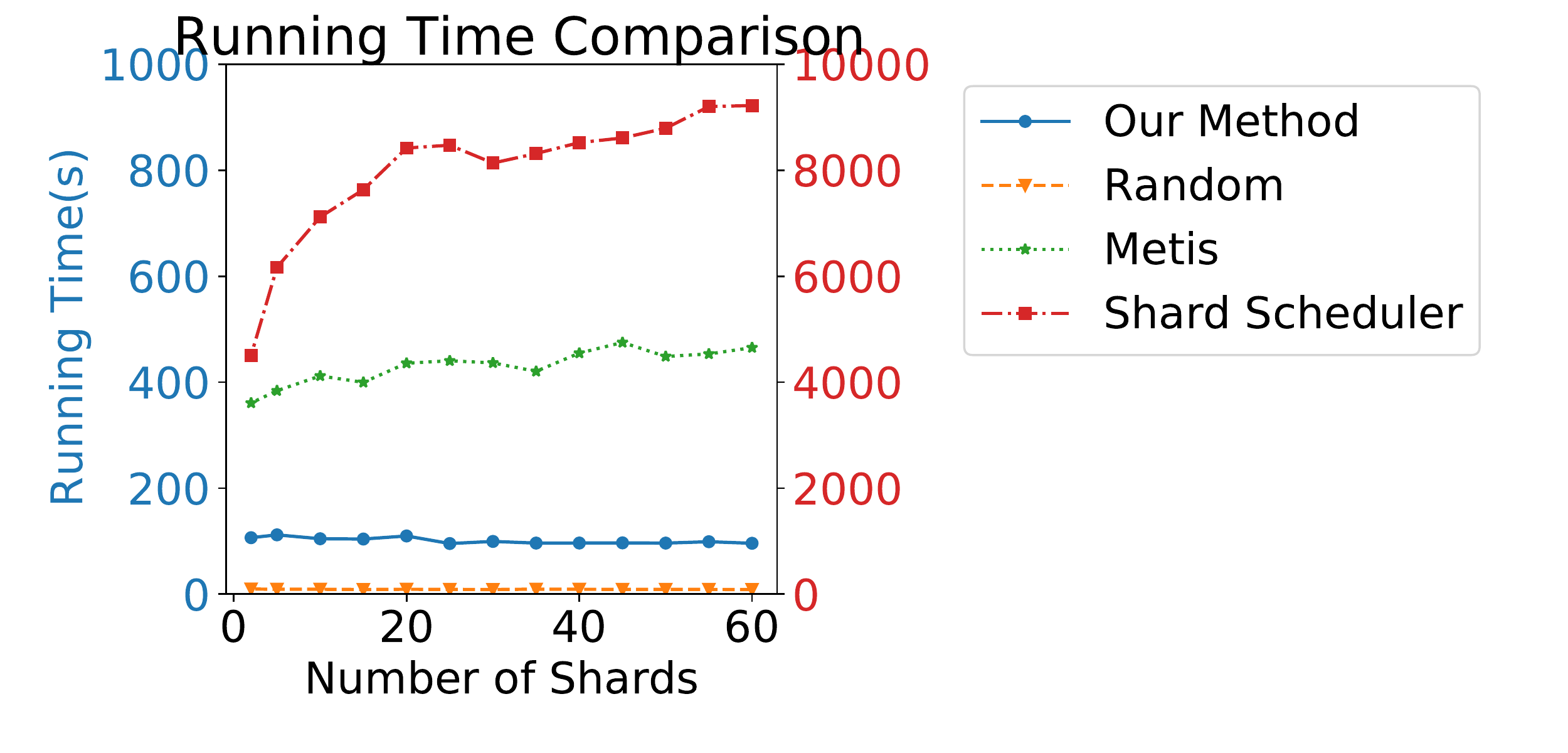}
      \caption{ $\Difficulty = 10$}
    \end{subfigure}
    \caption{Running time comparison with various $\numrelashards$ and $\Difficulty$.  The y-axis in the right-hand-side is for the transaction-level method Shard Scheduler in different scale.}
    \label{global_time}
\end{figure*}

\begin{figure*}[ht]
\captionsetup[subfigure]{justification=centering}
    \centering
    \begin{subfigure}{0.74\textwidth}
    \includegraphics[width=0.99\textwidth]{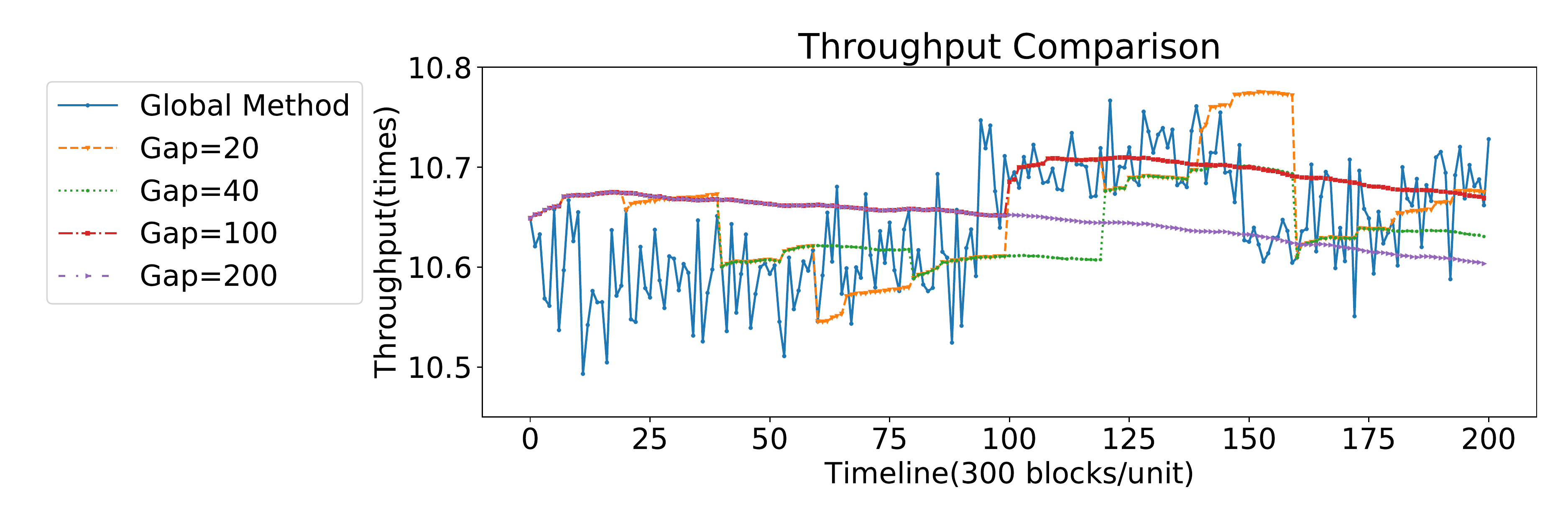}
    \caption{Throughput Comparison}
    \label{evo_throughput}
    \end{subfigure}
    \begin{subfigure}{0.24\textwidth}
     \centering
    \includegraphics[width=0.99\textwidth]{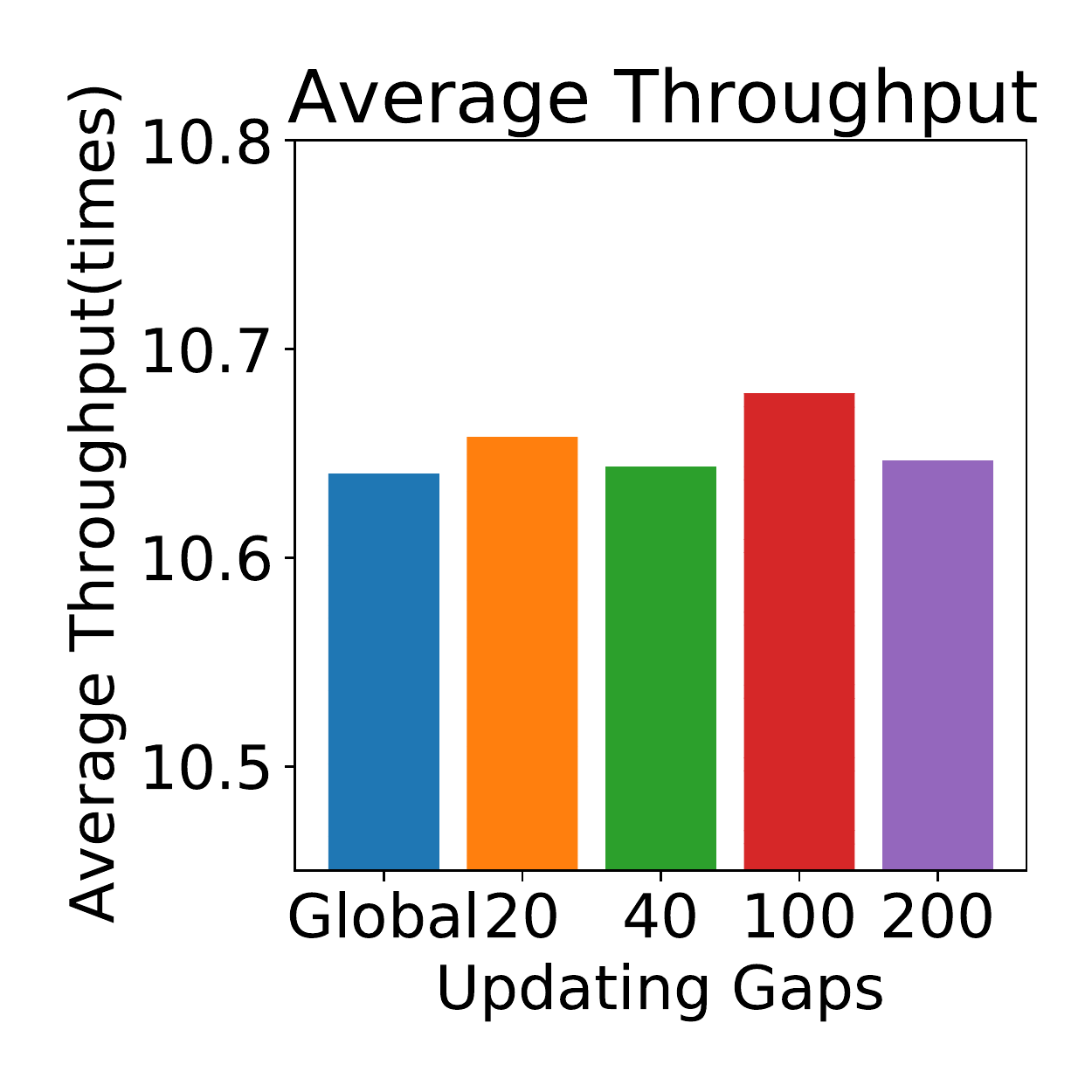}
    \caption{Throughput in Average}
    \label{evo_Average}
    \end{subfigure}
    \caption{The throughput comparison with $\updateada =300$ blocks and various global updating gap $\updateglo$. ~\ref{evo_throughput} shows the detailed evolution patterns and ~\ref{evo_Average} shows the average throughput of~\ref{evo_throughput}.}
    \label{Incre_throughput}
\end{figure*}

\begin{figure}[ht]
     \centering
     \includegraphics[width=0.48\textwidth]{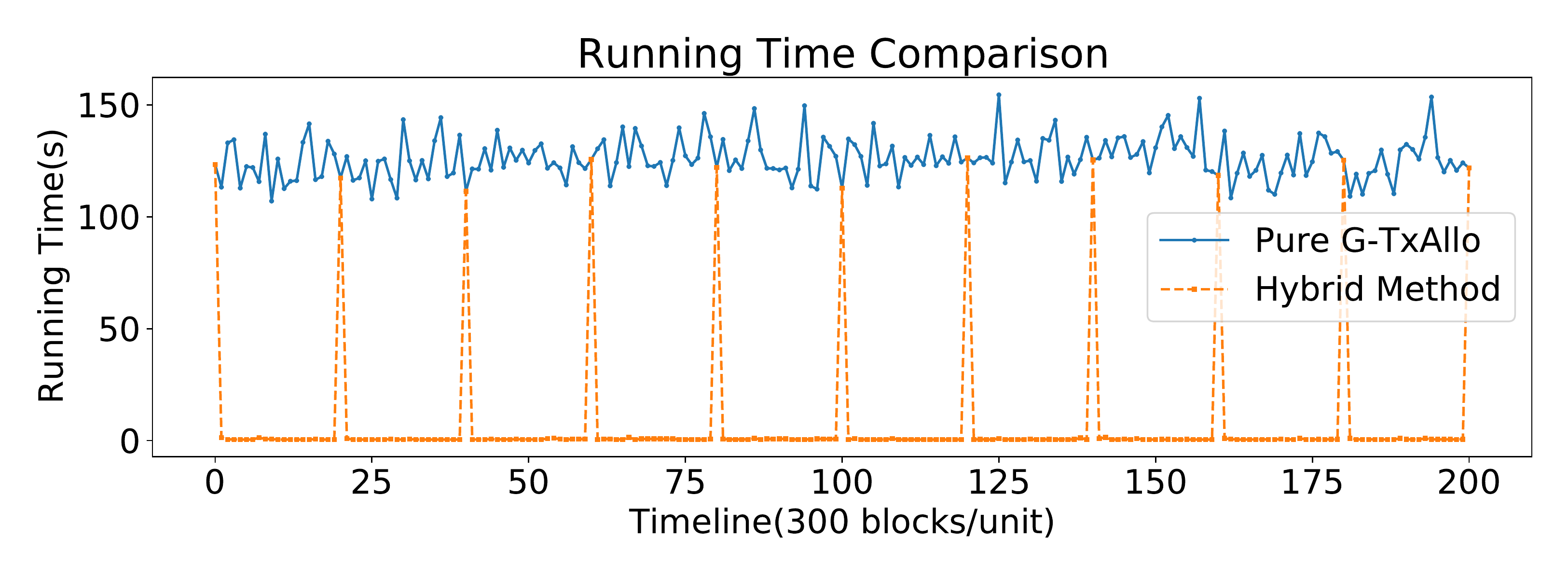}
     \caption{A case study example of running time comparison for pure \globalNAME{} and hybrid \NAME{} with $\updateada =300$ and $\updateglo =20 *300$ blocks.}
     \label{Incre_time}
\end{figure}

\subsubsection{Transaction Confirmation Latency Comparison}

Transaction confirmation latency is another important factor for a better user experience. The confirmation latency is different among shards. For transactions in over-loaded shards, it takes longer latency for the commitment. Figure~\ref{global_average_latency} and Figure~\ref{global_worst-case_latency} compare the average latency and the worst-case latency in the most over-loaded shard, respectively. 

 Figure~\ref{global_average_latency} shows that \globalNAME{} achieves the best average confirmation latency with every $\Difficulty$ and $\numshard$. In most circumstances, the confirmation latency is less than two blocks, which means clients only need to wait for two blocks on average (roughly 25s in Ethereum) for their transactions to be committed. The confirmation latency for our method is superior to the baselines when $\Difficulty$ is large, thanks to the effectiveness of reducing the cross-shard transactions.

For the worst-case latency comparison in Figure~\ref{global_worst-case_latency}, Shard Scheduler achieves the best results as its allocation does not involve overloaded shards as detailed in Figure~\ref{workload distribution}.
\globalNAME{} is better than the other two baselines as it considers workload balance to avoid the over-loaded problem.

\subsubsection{Running Time Comparison}
\label{sec_Time_global}
Figure~\ref{global_time} compares the running time with baselines. 
To get the account-shard mapping of the total 12 million accounts, taking $\Difficulty=2$ as an example, the average running time of Shard Scheduler, METIS and \globalNAME{} is 3447.9 seconds, 422.7 seconds and 122.3 seconds (including initialization with 67.6 seconds),  respectively. 
Shard Scheduler is short on running time as it updates account-shard mapping when each transaction comes.
For graph-based methods, \globalNAME{} achieves more than two times more efficient than the METIS method.

\subsubsection{Conclusions}
 This section illustrates that \globalNAME{} achieves (1) the lowest cross-shard transaction ratio $\crossratio$, (2) the best workload balance except for the transaction-level method Shard Scheduler,  (3) the best throughput $\throughput$, (4) the lowest average transaction confirmation latency and the second position for the worst-case latency and (5) the fastest running time except for random allocation. 
 In addition, we illustrate that \globalNAME{} can achieve self-adjustable results for different $\Difficulty$.

\subsection{\adaNAME}
\label{sec_ada_results}
This section evaluates \adaNAME{} and compares the evolution of throughput and running time. \adaNAME{} aims to get approximated account-shard mapping based on the previous result and new transactions. We show that it does not sacrifice the performance much in terms of throughput $\throughput$, whereas the running time is much shorter.

For the settings, we partition the entire dataset by the ratio of $9:1$. \globalNAME{} is first run on the transactions from block 10,000,000 to block 10,540,000 to get the last-step allocation results and the remaining transactions from block 10,540,001 to block 10,600,000 are used for the evaluation of $\adaNAME{}$.

\subsubsection{Throughput Comparison}
With new blocks generating, Figure~\ref{Incre_throughput} shows the comparison of throughput evolution with different updating gap $\updateglo$. Each time step consists of 300 blocks, approximately one hour in Ethereum data. 
In each time step, $\adaNAME{}$ is run based on the newly-generated transactions in these 300 blocks as well as the allocation results in the last time step.
In the domain of the y-axis, from $10.45$ to $10.8$ times, the curve of \globalNAME{} fluctuates as the account-shard mapping is completely updated without adaptive information from previous results.

For \adaNAME{}, to show the performance loss with time, we conduct \globalNAME{} with different global updating gaps: $\updateglo/300 =$ 20, 40, 100, 200. 
 Taking the longest updating gap $\updateglo/300 = 200$ as an example, the system throughput of pure \adaNAME{}  decreases slowly as in Figure~\ref{evo_throughput}. 
 This indicates that even if \globalNAME{} is conducted as long as every nine days (the running time of \globalNAME{} is around 122 seconds in the last section), the throughput loss of pure \adaNAME{} is still acceptable. 
 Figure~\ref{evo_Average} in the right-hand-side shows the average throughput for different global updating gaps. There is no significant difference in the average throughput for various $\updateglo$ and even the transaction pattern fluctuation in the data could have more influence on the throughput.

\subsubsection{Running Time Comparison}

Figure~\ref{Incre_time} shows the running time comparison between pure \globalNAME{} and hybrid methods with \adaNAME. 
The global and adaptive updating gaps are set as 20 and 1 time steps for the case study example, respectively. It is impressive that \adaNAME{} only takes about 0.55 seconds for execution every hour. This is negligible compared to 122 seconds of using \globalNAME{} and 422 seconds of using the METIS method, as shown in Section~\ref{sec_Time_global}.
Such fast execution makes it more practical for real-world applications. For example, the average block generation time in Ethereum is between 12 to 14 seconds.
The allocation updating latency is reduced from about 30 times block generation time to only about 4\%, making it 800 times more efficient.

\section{Integration to Existing Sharding Protocols}
\label{sec_integration}
As discussed in Section~\ref{sec_background}, this work focuses on account-based permissionless blockchains, which can be categorized into two types:
\begin{enumerate}
 \item Fully replicated state with sharded transaction processing (e.g., Monoxide\footnote{To achieve desired security guarantee, Monoxide employs Chu-ko-nu mining, requiring knowledge of the full state of every shard.}~\cite{wang2019monoxide}, Elastico~\cite{luu2016secure} and Zilliqa~\cite{zilliqa})
 \item Sharded state and transaction processing. (e.g., Omniledger~\cite{kokoris2018omniledger}, RapidChain~\cite{zamani2018rapidchain}, Ethereum 2.0~\cite{upgrades}, and Chainspace~\cite{al2017chainspace})
\end{enumerate}

For type 1 systems, \NAME{} can be applied directly as the underlying system already requires the knowledge of all states in all shards.
For type 2 systems, \NAME{} may introduce additional storage overhead but not network communication overhead. In particular, these systems require periodic re-shuffling to relocate miners to different shards~\cite{wang2019sok}. This is required due to the problem of node churning and single-shard take-over attacks. During the re-shuffling process, miners must retrieve the state of the new shard to which the miner is reallocated. In practice, retrieving the state requires information dissemination through the peer-to-peer network of the permissionless blockchain. Thus, every miner eventually receives all state information of all shards. While in type 2 sharded blockchains, miners may only keep non-related state information in the cache to forward it to other peer-to-peer network nodes. In our setting, they are required to store these data. Note that the stored data may not be the latest, but it is still enough for our algorithm.
Thus, our proposal may require additional storage but not additional network communications. 
Therefore, in both types of sharded systems, we do not require additional operations for data migration, as all states of all shards are eventually known to miners, which implies a virtual global ledger.

\section{Conclusions}\label{sec_conclu}
This work investigated the transaction allocation problem in sharded blockchains to reduce the expensive cross-shard transactions. We converted the key concepts from the blockchain sharding to a transaction graph and proposed \NAME{} to optimize the throughput on this graph. The experiments illustrate that the cross-shard transactions can be significantly reduced from 98\% to 12\%, and the execution time is approximately 800 times faster. In the meantime, the workload balance is also well maintained in different shards. For the adaptive updating method \adaNAME{}, the performance loss is acceptable even when being continuously conducted for a long period of 9 days. As this work and existing works rely on the assumption that future transaction patterns are similar to historical transactions, we leave the prediction of future transactions as our future work.

\bibliographystyle{IEEEtran}
\bibliography{MyRefs}

\end{document}